\documentclass[%
 aip,
 jap,
rsi,%
 amsmath,amssymb,
 reprint,%
]{revtex4-1}

\usepackage{graphicx}% Include figure files
\usepackage{dcolumn}% Align table columns on decimal point
\usepackage{hyperref}
\usepackage{bm}% bold math
\usepackage{xcolor}

\usepackage{tabularx}
\newcolumntype{Y}{>{\centering\arraybackslash}X}
\begin{document}

\preprint{AIP/123-QED}

\title{Thermal Expansion in Insulating Solids From First Principles}% Force line breaks with \\

\author{Ethan T. Ritz}
\affiliation{ 
Sibley School of Mechanical and Aerospace Engineering, Cornell University, Ithaca NY 14853, USA}%
\author{Sabrina J. Li}
\author{Nicole A. Benedek}
\email{nbenedek@cornell.edu}
\affiliation{%
Department of Materials Science and Engineering, Cornell University, Ithaca NY 14853, USA
}%

\date{\today}% It is always \today, today,

\begin{abstract}
In this Tutorial, we describe the use of the quasiharmonic approximation and first-principles density functional theory (DFT) to calculate and analyze the thermal expansion of insulating solids. We discuss the theory underlying the quasiharmonic approximation, and demonstrate its practical use within two common frameworks for calculating thermal expansion: the Helmholtz free energy framework and Gr\"uneisen theory. Using the example of silicon, we provide a guide for predicting how the lattice parameter changes as a function of temperature using DFT, including the calculation of phonon modes and phonon density of states, elastic constants, and specific heat. We also describe the calculation and interpretation of Gr\"uneisen parameters, as well as how they relate to coefficients of thermal expansion. The limitations of the quasiharmonic approximation are briefly touched on, as well as the comparison of theoretical results with experimental data. Finally, we use the example of ferroelectric PbTiO$_3$ to illustrate how the methods used can be adapted to study anisotropic systems.
\end{abstract}

\pacs{Valid PACS appear here}% PACS, the Physics and Astronomy
%Use showkeys class option if keyword
\maketitle

% \begin{quotation}
% The ``lead paragraph'' is encapsulated with the \LaTeX\ 
% \verb+quotation+ environment and is formatted as a single paragraph before the first section heading. 
% (The \verb+quotation+ environment reverts to its usual meaning after the first sectioning command.) 
% Note that numbered references are allowed in the lead paragraph.
% %
% The lead paragraph will only be found in an article being prepared for the journal \textit{Chaos}.
% \end{quotation}

\section{\label{sec:level1} Introduction}

\begin{figure}
\centering
\includegraphics[width=\columnwidth]{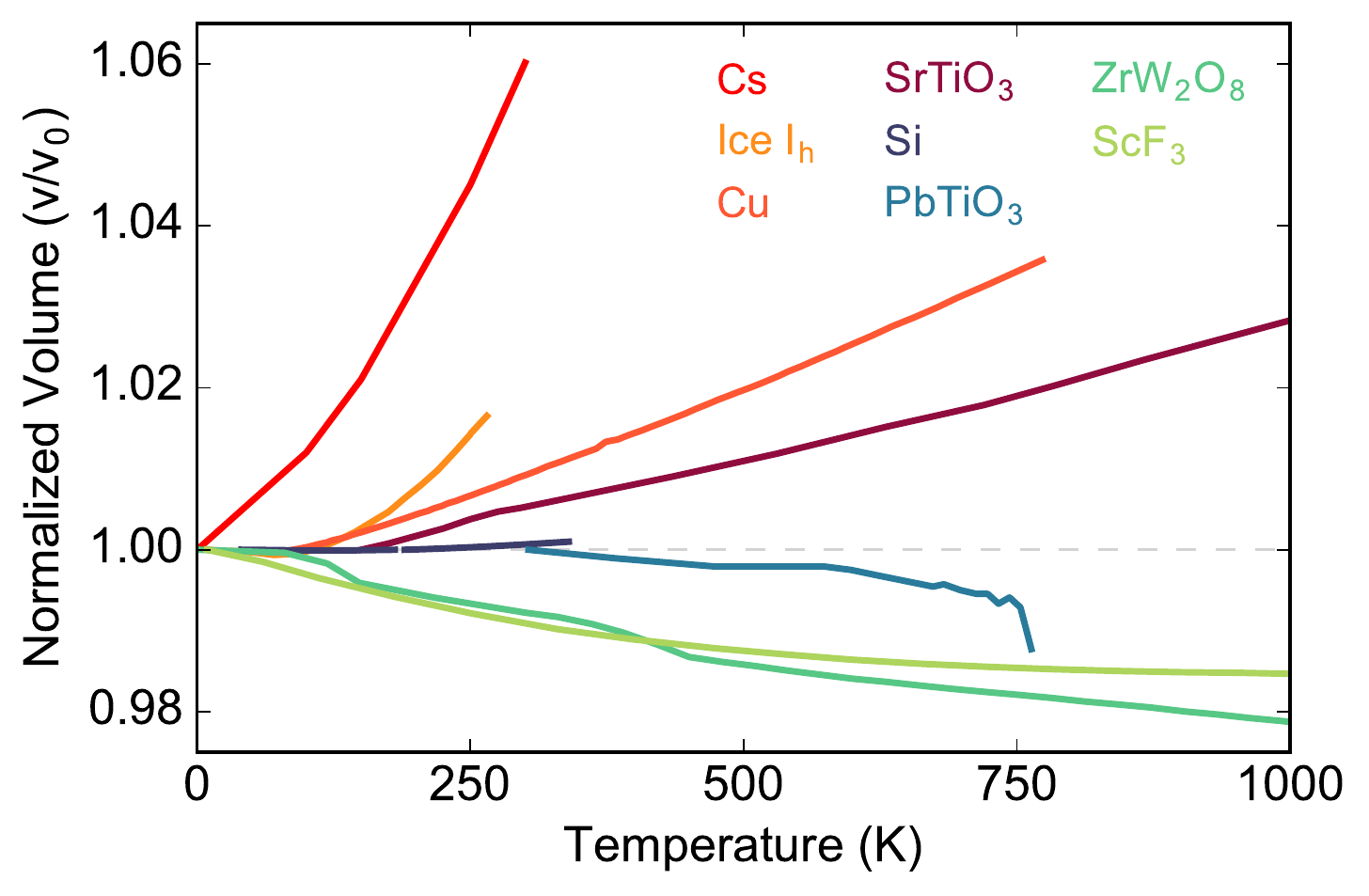}%{19_0628_TE_matl27.pdf}
\caption{Thermal strains in various positive and negative thermal expansion materials. \cite{anderson1969equation,rottger1994lattice,nix1941thermal,de1996high,lytle1964x,shah1972thermal,lyon1977linear,shirane1951phase,mary1996negative, greve2010pronounced}}
\label{fig:materials_thermal_expansion}
\end{figure}

Materials experience thermal strain --- changes in volume or shape --- as temperature changes. These changes are usually quite small, as Figure \ref{fig:materials_thermal_expansion} shows. For example, the lattice parameters of elemental cesium, the material with the largest coefficient of thermal expansion in the CRC Handbook,\cite{crc} change by less than 3\% over a temperature range of 100 K. However, although small, the presence of thermal strains can have profound implications for technological design. For instance, the strain produced by temperature changes can lead to the failure of components in semiconductor devices. Managing the deleterious effects of thermal strains is also critical to the safe operation of aircraft and spacecraft. From a fundamental perspective, understanding the relationship between thermal expansion, crystal structure, and chemical bonding is an area of significant current research in the solid-state physics and chemistry communities. In particular, elucidating the microscopic causes of the relatively rare phenomenon of negative thermal expansion (where instead of expanding with increasing temperature, a solid shrinks) is a major open challenge.

In the same manner as mechanical strain, thermal strain $\varepsilon^T_{ij}$ can be expressed as a 3 $\times$ 3 rank-two tensor, often as the product of the thermal expansion tensor, $\alpha^T_{ij}$, and temperature change $\Delta T$:
\begin{equation}
    \varepsilon_{ij}^T \equiv  \alpha^T_{ij}\Delta T.
    \label{thermTensor}
\end{equation}
The magnitude of $\alpha_{ij}^T$, commonly known as the coefficient of thermal expansion, describes the magnitude of expansion with temperature in the neighborhood of temperature $T$, while its sign indicates whether expansion is positive or negative. As it represents a ``strain per temperature", and since strain is dimensionless, it has units of $T^{-1}$.  The coefficient of volumetric thermal expansion, $\alpha^T_v$, which describes volumetric strain as a function of temperature through $\Delta V=\alpha^T_v\Delta T$, is equivalent to the trace of the thermal expansion tensor and also has units of $T^{-1}$. In a noncubic material, each unique crystallographic axis will generally have a different value (and possibly sign) of $\alpha_{ij}^T$, because crystals typically expand or contract at different rates with temperature along different directions. However, in a cubic material, since each crystal axis is constrained by symmetry to expand at the same rate, $\alpha_{ij}^T$ reduces to the diagonal matrix $\frac{\alpha_v^T}{3} \cdot \mathbf{1}$, where $\mathbf{1}$ is the identity matrix. The calculation of $\alpha_{ij}^T$ is usually a central aspect of theoretical studies of thermal expansion. Throughout the remainder of the Tutorial, we will drop the superscript $T$ denoting thermal strain and coefficients of thermal expansion, and all strain should be interpreted as thermal strain unless otherwise noted.

The goal of this Tutorial is to both briefly review a theoretical framework to understand thermal expansion, and to provide a worked example to calculate it using first-principles density functional theory (DFT) and the quasiharmonic approximation. We focus on insulating (and nonmagnetic) materials in which the main carriers of heat are phonons, and in which the effects of electron-phonon coupling on thermal transport processes can usually be ignored. We hope to provide the tools necessary for a reader familiar with both basic knowledge of solid-state physics and DFT to begin to simulate and analyze thermal expansion under the quasiharmonic approximation, as well as to assess the validity of their results. For a broader overview of historical background, theoretical framework, and state-of-the-art in (both positive and negative) thermal expansion studies, we refer the reader to the textbooks by Wallace (Ref. \citenum{wallace}) and Grimvall (Ref. \citenum{grimvall1999thermophysical}), as well as excellent review articles in Refs. \onlinecite{barron1980thermal,lind2012two,barrera2005negative,fultz2010vibrational,chen15,coates19,attfield2018mechanisms}.

\section{\label{Theory}Theory}
\subsection{\label{sec:Free Energy Framework} The Quasiharmonic Approximation}
Anharmonicities of the crystal lattice are the origin of thermal expansion in solid materials. There are typically two sources of anharmonicity that are relevant to thermal expansion: the coupling between phonons and strain (changes in the unit cell volume with temperature), and phonon-phonon coupling. In the quasiharmonic approximation, we assume that each phonon acts as an independent harmonic oscillator -- phonon-phonon coupling is ignored, and the energy of a single phonon mode does not depend on the occupation of any other phonon modes. In this approximation, the only contribution to the thermal expansion of a crystal comes from the coupling of phonons to changes in the unit cell dimensions. The quasiharmonic approximation breaks down for phases that are not dynamically stable at zero temperature (the temperature of a DFT simulation), since in these cases the phonon dispersion curve contains phonon modes with imaginary frequency. The quasiharmonic approximation also becomes increasingly poor as the temperature approaches the melting point of a given material because phonon-phonon coupling is strong at high temperatures.\cite{allen2015anharmonic,horton1974dynamical,wallace} Similar difficulties arise near structural phase transitions, the energy landscape of which can be dominated by higher-order phonon-phonon interactions or phonon mode instabilities that the quasiharmonic approximation does not account for. Many techniques have been developed for going beyond the quasiharmonic approximation and while this Tutorial will not discuss them in depth, we refer the reader to the relevant literature for further details. \cite{hellman2014phonon,kim2018nuclear,romero2015thermal,hellman2011lattice,ravichandran2018unified,tadano2015self,errea2013first,monserrat2013anharmonic,gillis1968properties,van2016anomalous,bianco2017second}

\subsection{Phonon Thermodynamics}
The equilibrium structural parameters $a_n$ of a given crystal structure at a given temperature $T$ are those that minimize the Helmholtz free energy $F(T,a_n)$ at that temperature (the number of unique $n$ depends on the crystal system in question). Note that in the most general case, $a_n$ can include lengths of the vectors defining the unit cell, their relative angles, \emph{and} internal degrees of freedom, such as the positions of nuclei with free parameters in their Wyckoff site. Our immediate objective is therefore to determine how to calculate $F(T,a_n)$ from first principles. For an isotropic system, such as a cubic crystal, there is a single lattice parameter $a$, which fully determines the shape of the unit cell.  $F(T,a)$ is defined \cite{wallace} as,

\begin{equation}
    F(T,a)\equiv-k_BT\mathrm{ln}(Z(T,a)),
\end{equation}

where $k_B$ is the Boltzmann constant, and $Z(T,a)$ is the canonical partition function given by, 
\begin{equation}
    Z(T,a)\equiv\sum_i e^{\frac{-E_i(T,a)}{k_BT}}.
    \label{partition}
\end{equation}

Here, $E_i(T,a)$ denotes the $i$'th vibrational energy level for a material with lattice parameter $a$ at temperature $T$. We have explicitly written out the functional dependence of both $F$ and $E_i$ on $T$ and $a$ to emphasize the link between these properties -- at a given $T$, each value of $a$ corresponds to a different set of $E_i(a)$, one of which will minimize $F(T,a)$. The problem of finding the lattice parameter as a function of temperature is thus equivalent to the problem of evaluating $E_i(a)$.

In an insulating system, phonons -- the normal mode solutions of the lattice-dynamic harmonic Hamiltonian -- are an appropriate basis through which to express the energy levels $E_i(a)$. Phonons are obtained by first expanding the crystal Hamiltonian as a function of the displacements of ions from their equilibrium positions --
\begin{widetext}
\begin{multline}
H=\Phi_0+\frac{1}{2}\sum_{N\nu} \sum_i M_{\nu} \dot{U}_i{{N}\choose{\nu}}^2+\frac{1}{2}\sum_{MN\mu\nu} \sum_{ij}\Phi_{ij}{{MN}\choose{\mu\nu}}U_i{{M}\choose{\mu}} U_j{{N}\choose{\nu}}\\ +\frac{1}{3!}\sum_{MNP\mu\nu\pi} \sum_{ijk}\Phi_{ijk}{{MNP}\choose{\mu\nu\pi}}U_i{{M}\choose{\mu}} U_j{{N}\choose{\nu}}U_k{{P}\choose{\pi}}+...
\label{hamilton}
\end{multline}
\end{widetext}
 Here, $M,N,$ and $P$ refer to unit cell indices, $\mu,\nu,$ and $\pi$ are atomic indices, and $i,j,$ and $k$ are Cartesian directions. $U_i{{M}\choose{\mu}}$ represents the displacement of atom $\mu$ in unit cell $M$ in direction $i$, and $\Phi_{ij}{{MN}\choose{\mu\nu}}$ is the second order force constant that couples displacements $U_i{{M}\choose{\mu}}$ and $U_j{{N}\choose{\nu}}$ to energy. The first term, $\Phi_0$, is the ground-state electronic energy of the system when all atomic displacements are zero, while the second is the kinetic energy due to nuclear motion. If we truncate the expansion at second order in the atomic displacements, then the eigenvalues of $H$ correspond to phonon modes with energy $\hbar\omega_{s,\mathbf{q}}$ where $\{s,\mathbf{q}\}$ are indices corresponding to phonon mode branch $s$ at wavevector $\mathbf{q}$. In the quasiharmonic approximation, the energy of the $i$'th vibrational mode does not depend on the occupation of any other modes. Thus, 
\begin{equation}
E_{i}(a)\approx\sum_{\omega_{s,\mathbf{q}}}(n^i_{s,\mathbf{q}}+\frac{1}{2})\hbar \omega_{s,\mathbf{q}},
\label{vib_energy}
\end{equation}
where $n^i_{s,\mathbf{q}}$ is the occupation number of mode $\{s,\mathbf{q}\}$ in energy level $i$. When substituted into Equation \ref{partition}, and using the properties of exponentials and logarithms, this approximation of the system energy levels results in the following expression for $F(a,T)$:

\begin{equation}
    F(T,a)=\Phi_0+\sum_{s,\mathbf{q}}\biggr\{\frac{{\hbar\omega_{s,\mathbf{q}}}}{2}+k_BT\mathrm{ln}\biggr(1-e^{-\frac{\hbar \omega_{s,\mathbf{q}}}{k_BT}}\biggr) \biggr\}\biggr|_a.
    \label{quasiharm}
\end{equation}
The sum over phonon modes can also be reformulated as an integral over the phonon density of states $g(\omega)$: \cite{kittel1976introduction}
\begin{multline}
    F(T,a)=\Phi_0+\\
    \int g(\omega)\biggr\{ \frac{\hbar\omega}{2}+k_BT\mathrm{ln}\biggr(1-e^{-\frac{\hbar \omega}{k_BT}}\biggr) \biggr\}\mathrm{d}\omega \biggr|_a
    \label{quasiharmDOS}
\end{multline}
As the only unknown on the right-hand side of Equation \ref{quasiharmDOS} is $g(\omega)$ evaluated for the system with lattice parameter $a$, a routine calculation of $\omega_{s,\mathbf{q}}$ on a sufficiently dense mesh of crystal momenta ($\mathbf{q}$) in reciprocal space is sufficient to evaluate $F(a,T)$ at all temperatures. 

The Helmholtz free energy provides all the information that is needed to obtain the lattice parameter as a function of temperature and hence, the thermal expansion coefficient. Using DFT, we can explicitly evaluate phonon frequencies for a range of $a$ values, then use them to calculate $F(T,a)$ for a given $T$. We can then find the value of $a$ which minimizes $F(T,a)$ -- this value of $a$ is the lattice parameter of the system at temperature $T$. Then, thermal expansion coefficients can be easily calculated from the slope of $a(T)$. However, the computational techniques and hardware powerful enough to do these calculations have only been available in the last couple of decades. A wealth of analytical strategies, such as the Gr\"uneisen framework, had previously been developed for exploring and calculating thermal expansion using the quasiharmonic approximation. Although these strategies are typically numerically cumbersome, they provide valuable physical insights into the origins of thermal expansion, as well as a bridge to understanding older thermal expansion literature. Below we present a worked example on how to calculate thermal expansion coefficients using both the free energy framework outlined above and Gr\"uneisen theory, described below. We additionally show how the results of thermal expansion calculations can be explored and analyzed using Gr\"uneisen theory.

\subsection{\label{sec:PhonAm} Gr\"uneisen Theory of Thermal Expansion}
The third, fourth and higher-order terms in Equation \ref{hamilton} can be used to describe both phonon-phonon coupling and phonon-strain coupling. Phonon-phonon coupling is essential to understanding and computing phonon scattering rates and thermal conductivity. As the form of Equation \ref{quasiharm} arises from the assumption that phonons do not couple to other phonons, we will not discuss this coupling in detail. Instead, the quasiharmonic approximation of the Helmholtz free energy relies on an assumption that phonon coupling to strain is the only non-negligible source of anharmonic behavior in the system. As the shape of the unit cell changes with temperature, the frequencies of the phonons can change as a result, due to coupling through third (and higher) order force constants to strain or volume changes. The third-order terms are closely related to the Gr\"uneisen parameters of the system, and can be used to evaluate them in closed-form \cite{horton1974dynamical,fabian1997thermal}. 

Mode Gr\"uneisen parameters are intrinsic material properties describing the derivative of the frequency of a single phonon mode $\omega_{s,\mathbf{q}}$ with respect to some degree of freedom (for example, volume, strain, electronic or spin degrees of freedom \cite{grimvall1999thermophysical}). There are both many choices of degrees of freedom and ways to compute the corresponding derivatives, so there are many ways in which the mode Gr\"uneisen parameter can be defined. A common definition in the thermal expansion literature, the volumetric Gr\"uneisen parameter, involves a simple derivative of phonon frequency with respect to volume:\cite{munn1975gruneisen, ashcroft1976solid, dove1993introduction,kaviany2014heat}
\begin{equation}
    \gamma^V_{s,\mathbf{q}}\equiv-\frac{\partial \ln \omega_{s,\mathbf{q}}}{\partial \ln V}=-\frac{V}{\omega_{s,\mathbf{q}}}\frac{\partial \omega_{s,\mathbf{q}}}{\partial V}.
    \label{grun_V}
\end{equation}

 In cubic systems, there is a unique value of $\gamma^V_{s,\mathbf{q}}$ for each phonon mode, as the configuration change associated with a change in volume is exactly defined by symmetry. In this context, a configuration is the set of parameters needed to uniquely define the unit cell of a particular Bravais lattice, that is, the three edge lengths $a$, $b$ and $c$ and the angles $\alpha$, $\beta$ and $\gamma$. In a noncubic material, there are many different configurations associated with each volume -- in fact, there are an infinite number. For example, in a tetragonal system, a single volume change could be achieved through an elongation or contraction of the $c$ axis, or alternatively, through an elongation or contraction of the $a$ axes. The computed values of $\gamma^V_{s,\mathbf{q}}$ for a noncubic system are therefore not unique and their calculation in these cases is of questionable value.

There also exist alternative definitions of the mode Gr\"uneisen parameter through phonon derivatives with respect to thermodynamic quantities other than volume. For example, taking the derivative of $\omega_{s,\mathbf{q}}$ along a pressure-volume isotherm yields:\cite{grimvall1999thermophysical,wallace}
%\begin{equation}
%    \gamma^G_{s,\mathbf{q}}\equiv-\biggr(\frac{\partial \mathrm{ln} \omega_{s,\mathbf{q}}}{\partial \ln V}\biggr)_T=B\biggr(\frac{\partial \ln \omega_{s,\mathbf{q}}}{\partial P}\biggr)_T=\frac{V}{c_{s,\mathbf{q}}}\biggr(\frac{\partial S_{s,\mathbf{q}}}{\partial V}\biggr)_T.
%    \label{grun_G}
%\end{equation}
\begin{equation}
    \gamma^G_{s,\mathbf{q}}\equiv-\biggr(\frac{\partial \mathrm{ln} \omega_{s,\mathbf{q}}}{\partial \ln V}\biggr)_T=B\biggr(\frac{\partial \ln \omega_{s,\mathbf{q}}}{\partial P}\biggr)_T.
    \label{grun_G}
\end{equation}
Here, $B$ is the bulk modulus and we use the superscript $G$ to denote the thermodynamic Gr\"uneisen parameter as in Ref. \citenum{grimvall1999thermophysical}. Whereas $\gamma^V_{s,\mathbf{q}}$ is only uniquely defined for cubic systems, Equation \ref{grun_G} is uniquely defined for both cubic and anisotropic systems, and reduces to $\gamma^V_{s,\mathbf{q}}$ in the cubic case. Most experimental techniques for finding volumetric Gr\"uneisen parameters actually report thermodynamic mode Gr\"uneisen parameters, since volume changes are usually practically accomplished with application of pressure. For isotropic systems, calculated and experimentally measured Gr\"uneisen parameters are directly comparable, since, again, in these cases Equations \ref{grun_V} and \ref{grun_G} are equivalent.

Finally, the generalized mode Gr\"uneisen parameter \cite{leibfried1961solid,wallace,munn1975gruneisen,collins1964chapter,horton1974dynamical,barron1967analysis} is defined as the derivative of phonon frequency with respect to an infinitesimal strain $\varepsilon_{ij}$:
\begin{equation} \label{eq: grun_aniso}
    \gamma^{ij}_{s,\mathbf{q}}\equiv -\frac{1}{\omega_{s,\mathbf{q}}}\frac{\partial \omega_{s,\mathbf{q}}}{\partial \varepsilon_{ij}}.
\end{equation}
Equation \ref{eq: grun_aniso} can accommodate a general distortion of the crystal lattice and can be calculated by simply applying strain along the relevant strain degrees of freedom and computing a numerical derivative of each $\omega_{s,\mathbf{q}}$. Note that Equation \ref{eq: grun_aniso} is \emph{not} equivalent to Equation \ref{grun_G} in the case of a noncubic material, and so care should be exercised in comparing the experimentally measured and theoretically calculated Gr\"uneisen parameters for anisotropic systems (Equation \ref{grun_G} can, of course, be calculated for a noncubic material, but it is far more computationally intensive and much less straightforward than calculating Equation \ref{eq: grun_aniso}). 

If the Gr\"uneisen parameters are known from experiments or have been calculated, they can be combined with Equation \ref{quasiharm} to approximate the coefficient of thermal expansion to first order in the strain-phonon coupling. The proof is well-known, and a detailed treatment for the isotropic case can be found in Refs. \onlinecite{ashcroft1976solid} and \onlinecite{grimvall1999thermophysical}. Since the thermal strains are constrained by symmetry to be isotropic as well, the thermal expansion tensor of Equation \ref{thermTensor} reduces to a scalar volumetric coefficient of thermal expansion $\alpha_v$, which fully defines the change in unit cell volume as a function of temperature. Here, 

\begin{equation}
\alpha_v=\mathrm{Tr}(\alpha^{ij})=\frac{1}{V}\biggr(\frac{\partial V}{\partial T}\biggr)_P.
\end{equation}

If the bulk modulus at temperature $T$ is defined as
\begin{equation}
    B_T=-V\biggr(\frac{\partial P}{ \partial V }\biggr)_{T},
    \label{bulkMod}
\end{equation}
\noindent
then the product of $\alpha_v$ and $B_T$ is given by,

\begin{align}
\begin{split}
    \alpha_vB_T=-\biggr(\frac{\partial P}{ \partial V }\biggr)_{T}\biggr(\frac{\partial V}{\partial T}\biggr)_P \\
    =\biggr(\frac{\partial P}{\partial T}\biggr)_V=\biggr(\frac{\partial S}{\partial V}\biggr)_T,
\end{split}
\end{align}
\noindent
where the last step involves a Maxwell relation. Using the entropic term of Equation \ref{quasiharm} for $S$,

\begin{align}
\begin{split}
     \alpha_vB_T=\frac{\partial}{\partial V}\biggr\{ \sum_{s,\mathbf{q}} k_B\mathrm{ln}\biggr(1-e^{-\frac{\hbar \omega_{s,\mathbf{q}}}{k_BT}}\biggr) \biggr\}\biggr|_{a,T}\\
= \sum_{s,\mathbf{q}} \frac{\hbar^2}{k_BT^2}\frac{\omega_{s,\mathbf{q}}}{n^2_{s,\mathbf{q}}}\frac{\partial \omega_{s,\mathbf{q}}}{\partial V}\biggr|_{a,T}\\
=     \sum_{s,\mathbf{q}}\frac{c_{s,\mathbf{q}}}{\omega_{s,\mathbf{q}}}\frac{\partial \omega_{s,\mathbf{q}}}{\partial V}\biggr|_{a,T},
   \label{dVol}
 \end{split}
\end{align}
\noindent
where $n_{s,\mathbf{q}}$ is the equilibrium occupation number of phonon modes with energy $\hbar\omega_{s,\mathbf{q}}$, and $c_{s,\mathbf{q}}$ is the mode specific heat, defined as the contribution to bulk specific heat $C^V$ of phonon modes with energy $\hbar\omega_{s,\mathbf{q}}$. Using the definition of the volumetric Gr\"uneisen parameter in Equation \ref{grun_V},

\begin{equation}
     \alpha_vB_T =\frac{1}{V}\sum_{s,\mathbf{q}}c_{s,\mathbf{q}}\gamma^V_{s,\mathbf{q}}\biggr|_{a,T}.
\end{equation}
\noindent
Then, defining the bulk or mean Gr\"uneisen parameter as
\begin{equation} \label{eq: bulk_gru}
    \gamma_{bulk}^{V} \equiv \frac{\sum_{s,\mathbf{q}}\gamma^{V}_{s,\mathbf{q}}c_{s,\mathbf{q}}}{\sum_{s,\mathbf{q}}c_{s,\mathbf{q}}},
\end{equation}
\noindent
we can express the coefficient of volumetric thermal expansion as, 

\begin{equation}
     \alpha_v =\frac{\gamma_{bulk}^{V}C^V}{B_TV}\biggr|_{a,T}.
     \label{alpha_vol}
\end{equation}

A close examination of the above proof provides many important insights. First, as all of $B_T$, $C^V$, and $V$ must be positive quantities in a stable system, the expression in Equation \ref{alpha_vol} implies that the sign of $\alpha_v$ in isotropic systems must be the same as the sign of $\gamma_{bulk}^{V}$. This well-known observation underpins the association of negative thermal expansion (negative $\alpha_v$) with negative bulk Gr\"uneisen parameters.

Next, the simple form of the coefficient of volumetric thermal expansion in Equation \ref{alpha_vol} is a direct result of the quasiharmonic approximation. This is a subtle but important point. The expression for the system vibrational energy levels in Equation \ref{vib_energy} contains the occupation number for only a single phonon mode. If we had included the effects of phonon-phonon coupling, then the derivative of entropy with respect to volume in Equation \ref{dVol} would involve products of different phonon modes and their average occupations. The Gr\"uneisen parameter expressions in Equations \ref{grun_V}--\ref{eq: grun_aniso} are measurable physical quantities regardless of which terms we include in the free energy, but it is the quasiharmonic approximation that allows for a relatively simple and straightforward relationship between the thermal expansion coefficient and Gr\"uneisen parameters. The practical consequence of this is that Equation \ref{alpha_vol} should not be used to calculate thermal expansion in a system for which the quasiharmonic approximation is not justified.

Lastly, as all the quantities on the right hand side of Equation \ref{alpha_vol} are themselves dependent on volume and temperature, they will change as the crystal undergoes thermal strain. Hence, a given value of $\alpha_v$ is only valid for a finite region around the $\{a_0,T_0\}$ for which it was calculated. As the system undergoes significant thermal strain away from $a_0$, the expression for $\alpha_v$ derived using the Gr\"uneisen framework will lose accuracy due to its failure to account for fourth and higher-order terms that couple phonon frequency to strain.

\section{\label{sec:Example} Worked Example: Silicon}
\begin{figure}
    \centering
    \includegraphics[width=8cm]{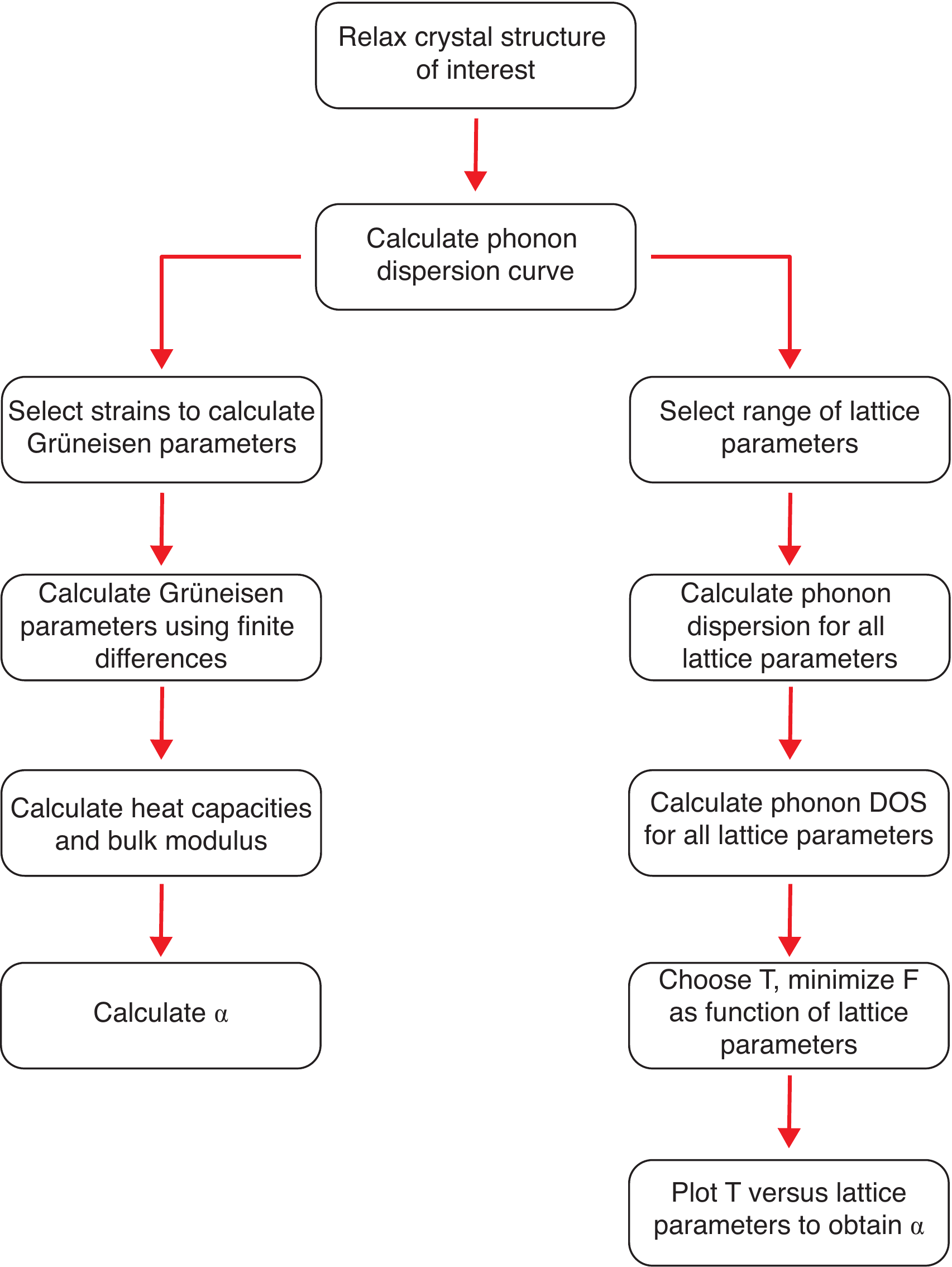}%{workflow.eps}
    \caption{Workflow illustrating the calculations required and the steps involved in calculating the thermal expansion coefficient from first-principles DFT from Gr\"uneisen theory (left branch) and from a full minimization of the Helmholtz free energy (right branch).}
    \label{fig:workflow}
\end{figure}

We now present a worked example demonstrating how to calculate the thermal expansion coefficient of silicon from first-principles DFT. Figure \ref{fig:workflow} shows the various steps involved in calculating $\alpha_v$ using both the Gr\"uneisen framework (left branch) and the free energy framework (right branch). Some of the steps shown in Figure \ref{fig:workflow} are also required for a calculation of thermal conductivity, \textit{e.g.} phonon dispersion curves. A recent excellent Tutorial\cite{mcgaughey2019phonon} describing the computational framework for predicting phonon properties and thermal conductivity from first principles provides a comprehensive overview of available DFT codes and their capabilities, the theory of lattice dynamics, different techniques for calculating phonon dispersion curves, and many other topics. Rather than repeat that information here, we focus on the specific calculations required for studying thermal expansion, and refer the interested reader to Ref. \onlinecite{mcgaughey2019phonon} for further details where indicated. We also note that thermal expansion can also be calculated using molecular dynamics platforms such as LAMMPS \cite{plimpton1995fast} with empirical potentials, rather than DFT. \cite{bouessel2014thermal,mcgaughey2014predicting,subramaniyan2008continuum} These techniques generally trade quantitative accuracy for computational tractability, and though we will not discuss these techniques further, they may prove to be a better (more practical) choice than DFT for systems with large unit cells or disorder.

Regardless of how the thermal expansion coefficient is calculated, the first step is to fully relax the crystal structure (lattice parameters and internal atomic coordinates) of interest. All the relaxations described below are 0 K calculations. The calculations were performed with DFT, as implemented in Quantum Espresso.\cite{giannozzi2009quantum} We used the Perdew-Burke-Ernzehof revised for solids (PBEsol)\cite{perdew2008restoring} exchange-correlation functional, with Garrity-Bennett-Rabe-Vandberbilit ultrasoft pseudopotentials.\cite{garrity2014pseudopotentials} The results of DFT calculations, such as structural parameters, elastic constants, and phonon frequencies can be highly dependent on the choice of exchange-correlation functional and pseudopotential. A well known example is provided by cubic SrTiO$_3$. In this material, calculating the volume using the local density approximation results in all phonon modes being stable at the zone center. However, calculating the volume using PBE will result in some modes having imaginary frequencies at the zone center. Hence, care must be taken to ensure that the choice of functional and pseudopotential is appropriate for the material under study -- for further details, again see the discussion in Ref. \citenum{mcgaughey2019phonon}. 

The lattice parameter was converged to within 0.001 \AA~with a 10$\times$10$\times$10 Monkhorst-Pack $\mathbf{k}$-point mesh and a plane wave energy cutoff of 50 Ry, compared to a 12$\times$12$\times$12 mesh and plane wave cutoffs up to 80 Ry. The forces on the atoms are zero by symmetry, however we relaxed the lattice parameters to nominally zero pressure. All lattice dynamical properties were calculated with density functional perturbation theory (DFPT)\cite{baroni2001phonons} on an 8$\times$8$\times$8 $\mathbf{q}$-point grid.

The phonon properties of a material depend sensitively on the crystal structure, so the forces on the atoms (when they are not zero by symmetry, as in this case) must be made as small as possible to ensure accurate phonon properties. A phonon dispersion curve should then be calculated for the relaxed structure to make sure that it is mechanically stable, that is, the phonon dispersion curve should not contain any modes with imaginary frequencies (in which case, the quasiharmonic approximation is not valid).

\begin{figure}
    \centering
    \includegraphics[width=\columnwidth]{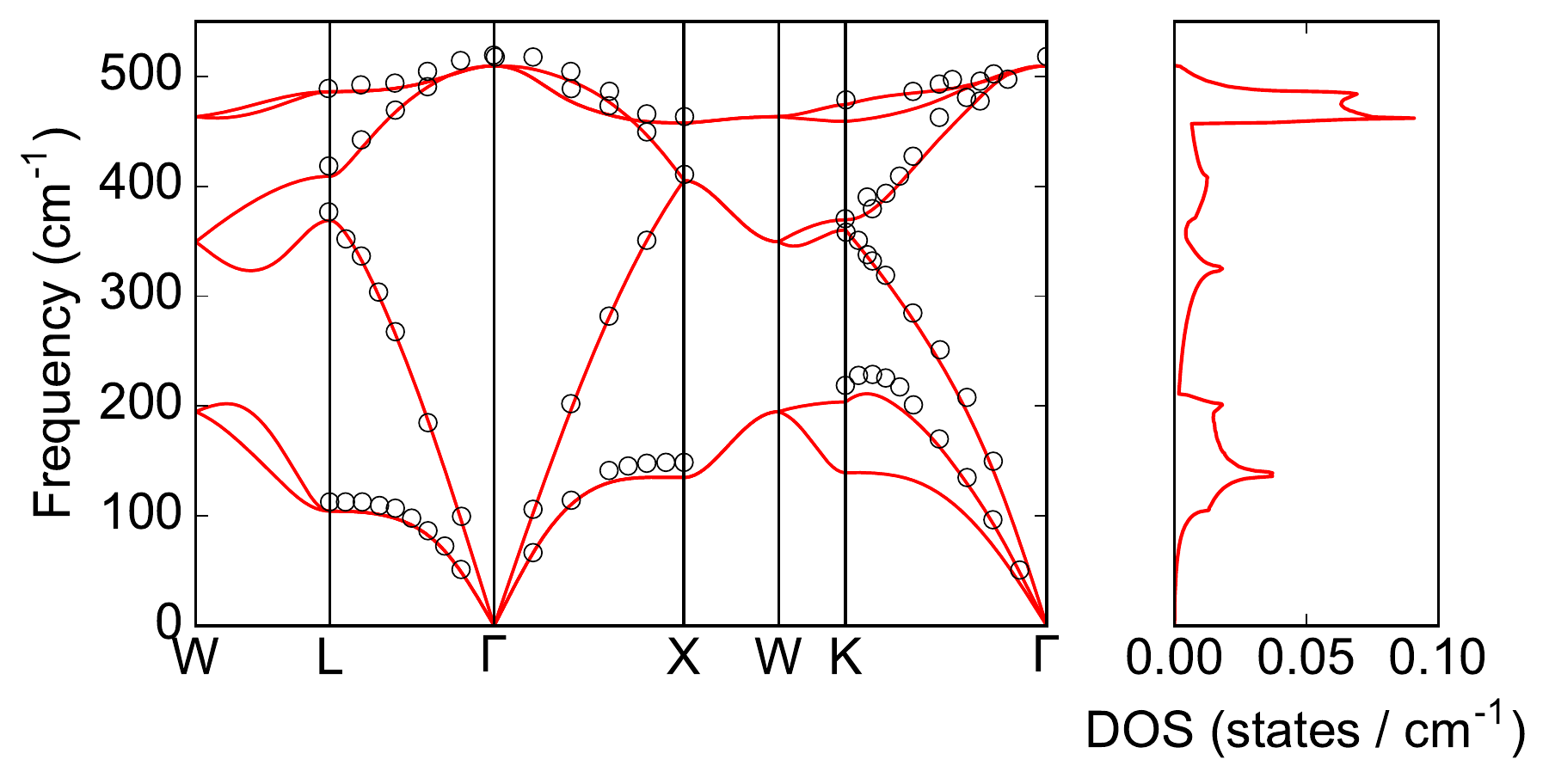}
    \caption{Silicon phonon dispersion curves and density of states from DFPT using the PBEsol functional at 0 K (in red), and from inelastic neutron scattering at 296 K (circles)\cite{dolling1963inelastic}. The high-symmetry points are in (conventional) Cartesian coordinates: W = (2$\pi/a$)(0.5, 1, 0), L = (2$\pi/a$)(0.5, 0.5, 0.5), $\Gamma$ = (0, 0, 0), X = (2$\pi/a$)(0, 1, 0), K = (2$\pi/a$)(0.75, 0.75, 0), where $a$ is the lattice parameter, 5.4319 \AA. See Section \ref{sec:Aniso} for a discussion on finite temperature calculations.}     
    \label{fig:Si_phonon_disp}
\end{figure}

Table \ref{tab:relaxed_lattice_comparison} shows the optimized lattice parameter of silicon from our DFT calculations; the agreement with experimental data is excellent. Figure \ref{fig:Si_phonon_disp} shows the phonon dispersion curve and phonon density of states (DOS) for our optimized silicon structure from DFPT calculations. As expected, there are no instabilities (phonon modes with imaginary frequencies). Note that phonon dispersion curves are usually calculated using either DFPT, or the method of supercells and finite displacements. The advantages and disadvantages of each technique will be discussed in Section \ref{sec:phonondisp_qha}.

\begin{table}
\caption{Optimized lattice parameter of silicon from our first-principles calculations compared with room temperature experimental data. The quasiharmonic calculations (indicated by QHA) were derived from the minimization of Helmholtz free energy as described in Section \ref{sec:TE_from_helm}. All DFT calculations used the PBEsol functional and Garrity-Bennett-Rabe-Vandberbilit ultrasoft
pseudopotentials.}
    \centering
    \begin{tabular}{l l r}
        \hline
         Method & Lattice parameter [\AA]  \\
         \hline
         DFT PBEsol 0 K (This work) & 5.4319\\
         DFT PBEsol QHA 0 K (This work) & 5.4414\\
         DFT PBEsol QHA 300 K (This work) &5.4423\\
         Neutron powder\cite{toebbens01} & 5.43053(7)\\
         X-ray powder\cite{ross14}& 5.4315(2)\\
         X-ray powder\cite{hom75}&5.430941(10)\\
         \hline
    \end{tabular}
    \label{tab:relaxed_lattice_comparison}
\end{table}{}

{}

\subsection{\label{sec:TE_from_helm} Thermal Expansion from Helmholtz Free Energy}

\subsubsection{\label{latRange}Select range of lattice parameters}

The first step required to calculate thermal expansion from a minimization of the Helmholtz free energy within the quasiharmonic approximation is to create a series of unit cells with varying lattice parameters. The lattice parameters that minimize $F$ at a given temperature are the equilibrium lattice parameters at that temperature. Care should be taken to ensure that the predicted equilibrium lattice parameters for each temperature fall within the range of lattice parameters used to compute $F$. Too small a range will result in predicted lattice parameters that erroneously ``hug" the maximum or minimum range, whereas too large a range wastes computational resources. If available, experimental thermal expansion data for the material under study or for a similar system can be used as a guide to select an appropriate range of strains -- if only $\alpha_v$ is reported, it can be multiplied by the temperature range of interest to generate a first approximation of what volumetric strains to expect. If such data is not available, a coarse mesh of $\pm$2-3\% strain centered around the calculated ground state lattice parameter should be a safe initial choice for most systems, and this can be iterated upon to find appropriate boundaries. Since silicon is cubic and the only possible symmetry-conserving strain is an isotropic volume change, this strain can be applied by simply varying the length of the single lattice parameter, $a$. For materials with free parameters in their atomic positions, unless the investigation explicitly includes them as degrees of freedom in the Helmholtz free energy optimization, the atomic positions must be fully relaxed for each set of strained lattice parameters. This will be discussed further in Section \ref{sec:Aniso}. 

\subsubsection{\label{sec:phonondisp_qha} Calculate phonon dispersion curve for all lattice parameters}

The calculation of phonon dispersion curves requires the harmonic terms $\Phi_{ij}$, or force constants, of the crystal Hamiltonian in Equation \ref{hamilton}. These are associated with the second-order terms of a Taylor expansion of potential energy in a basis of atomic displacements. The interatomic force constants are used to build the dynamical matrix $D$,

\begin{align}
\begin{split}
    D_{ij}(\mu\nu,\mathbf{q}) =  \frac{1}{N_0\sqrt{m_\mu m_\nu}}\sum_{MP}\Phi_{ij}{{MP}\choose{\mu \nu}}e^{i\mathbf{q}\cdot [\mathbf{R}(P\nu)-\mathbf{R}(N\mu)]}\\=\frac{1}{N_0\sqrt{m_\mu m_\nu}}\Phi_{ij}(\mu\nu,\mathbf{q}),
    \end{split}
\end{align}

where $i$ and $j$ again denote different Cartesian directions and $m_\mu$ and $m_\nu$ are respectively the masses of the $\mu$'th and $\nu$'th atoms, $M$ and $P$ label individual unit cells throughout the full sample containing $N_0$ unit cells, and $\mathbf{R}(P\nu)$ is the equilibrium position of atom $\nu$ in unit cell $P$. The eigenvalues of $D$ are the phonon frequencies and the eigenvectors represent the patterns of atomic displacements for each phonon mode. The calculation of interatomic force constants using DFT is now routine so we will not describe the technical details here except to note that there are generally two different approaches: DFPT\cite{baroni2001phonons} and the method of finite displacements using supercells. In DFPT, the change in ground state energy with respect to atomic displacements is found through a self-consistent linear response theory involving perturbations to wave functions, external potential, and electron density with respect to those atomic displacements. Phonon frequencies and eigenvectors can be obtained at arbitrary $\mathbf{q}$ using only the primitive unit cell of the material of interest. In contrast, the finite displacement method requires the construction of a supercell of the primitive unit cell. The forces induced by finite displacements of individual atoms are then computed using the Hellman-Feynman theorem. Each approach has advantages and disadvantages that may make them more or less appropriate in different contexts. The method of finite displacements can be used with any DFT code that can compute forces, whereas the DFPT algorithm must be specially implemented and is quite complex. However, in a finite displacement calculation the supercell needs to be made as large as practicable, since the number of allowed $\mathbf{q}$ we can explicitly solve for using a supercell is related to the number of primitive unit cells used to build it. That is, a 2$\times$2$\times$2 supercell will allow for the calculation of phonons with $\mathbf{q}$ = 0 and $\mathbf{q}=(\frac{1}{2},\frac{1}{2},\frac{1}{2})$, but not $\mathbf{q}=(\frac{1}{4},\frac{1}{4},\frac{1}{4})$, whereas a 4$\times$4$\times$4 supercell allows for the calculation of phonons at all three wave vectors. Though phonons at $\mathbf{q}$ points that are not explicitly calculated can be approximated through interpolation, the results become more accurate as more $\mathbf{q}$ points are explicitly included, and the supercell size effectively becomes a convergence parameter. Despite the relative simplicity of the finite displacements method, we have encountered problematic behavior related to the convergence of phonon frequencies with respect to the supercell size, namely, phonon frequencies did not converge before the supercell became impracticably large (also see the discussion by McGaughey and co-workers\cite{mcgaughey2019phonon}). In some cases, the phonon dispersion curve developed anomalous features (such as sudden dips in frequency at non-zero $\mathbf{q}$) as the supercell size was increased. Since the predicted thermal expansion coefficient depends on the quality of the phonon dispersion curves used to calculate it, we typically use DFPT to compute harmonic-level phonon properties, and in agreement with Ref. \onlinecite{mcgaughey2019phonon}, we also recommend this practice.

\subsubsection{\label{sec:DOS} Calculate phonon density of states for all lattice parameters}
The phonon density of states (per unit volume) $g(\omega)$ is given by, \cite{ashcroft1976solid}
\begin{multline}
    g(\omega)=\sum_{s}\int \frac{d\mathbf{q}}{(2\pi)^3}\delta(\omega-\omega_s(\mathbf{q}))=\frac{1}{N}\sum_{s,\mathbf{q}}\delta(\omega-\omega_{s,\mathbf{q}}),
    \label{DOS}
\end{multline}

where the integral is over the first Brillouin zone, and $N$ is the number of wave vectors $\mathbf{q}$ in the sum. Many software packages can now automatically calculate the phonon DOS. However, calculating the DOS manually from the phonon dispersion curve is a valuable pedagogical exercise and we describe a simple procedure here.

The calculation of the phonon DOS essentially involves building a $\mathbf{q}$-averaged histogram of the phonon dispersion curve as a function of frequency. Note that a high-quality DOS calculation typically requires that the phonon dispersion curve be calculated on a very dense grid of $\mathbf{q}$-points (around 3-5 times denser than that required for good convergence of phonon frequencies, in our experience). We first select a DOS resolution, $\Delta \omega$, usually on the order of 1 cm$^{-1}$, and determine the upper bound on which the DOS is defined, $\omega_{max}$, which must be larger than the maximum phonon frequency in the phonon dispersion. These values are then used to define a discrete set of phonon frequencies $\Omega[n]$, or bins, where $n$ is an integer such that $0\leq n \leq \frac{\omega_{max}}{\Delta \omega}$ (with the quotient rounded up to the nearest integer):

\begin{figure}
    \centering
    \includegraphics[width = .93\columnwidth]{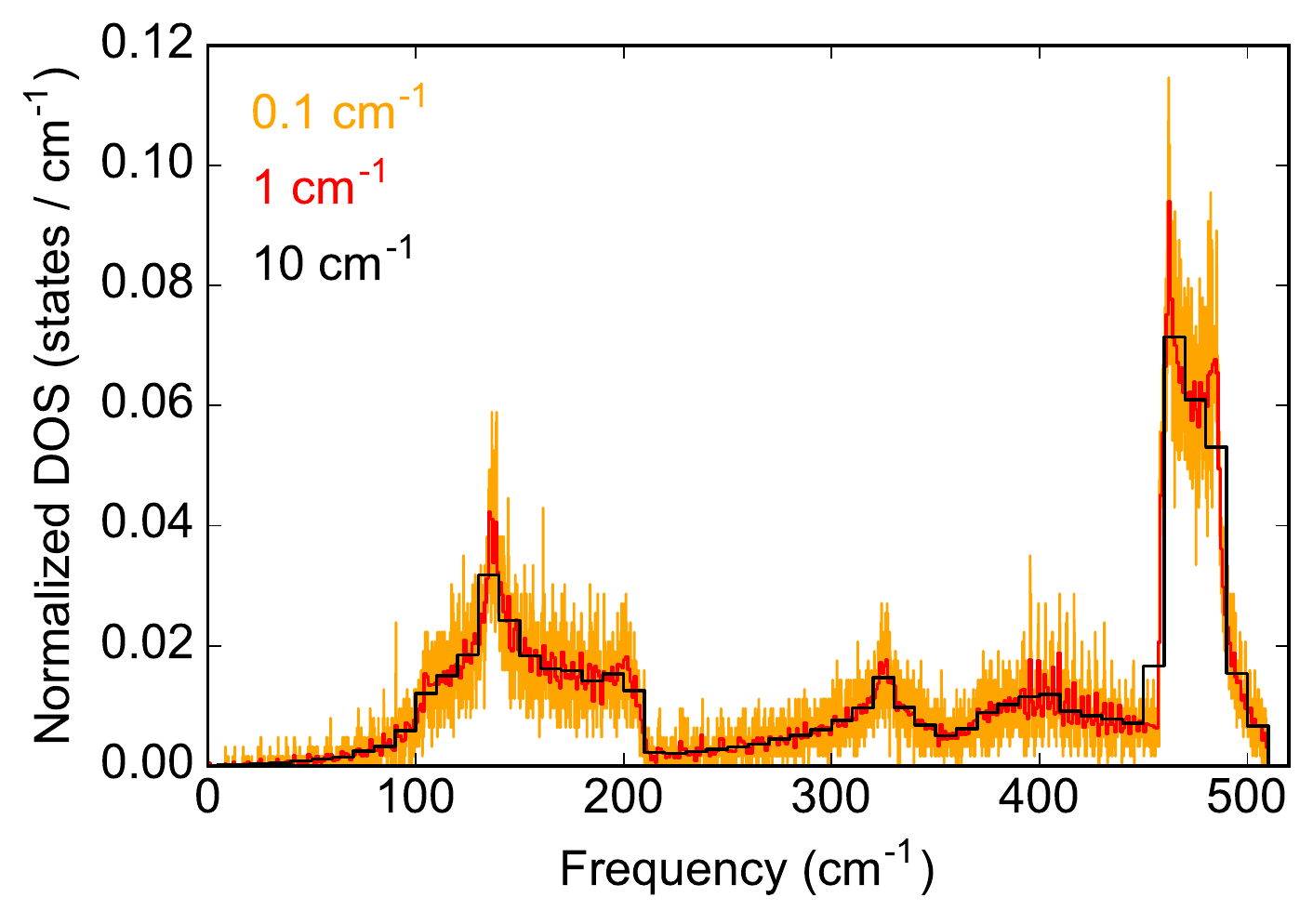}%{19_0807Si_dos_ork.pdf}
    \caption{Comparison of bin widths for calculation of the phonon density of states of silicon from DFPT using the PBEsol functional. Small bin widths result in a noisy density of states whereas large bin widths result in a loss of detail. The DOS here has been normalized by dividing by the number of sampled q-points, and dividing by the bin width. A sum of the normalized DOS gives six (three times the number of atoms in the unit cell), which is the number of phonon branches in the Si primitive unit cell.}
    \label{fig:dos_binwidth}
\end{figure}{}

\begin{equation}
    \Omega[n]\equiv n\Delta\omega.
\end{equation}

If the bins are too wide, important features will be averaged out, as shown in Figure \ref{fig:dos_binwidth}. However, if the bins are too narrow, the DOS becomes noisy. The binned frequencies are used to define a discrete DOS such that,
\begin{equation}
    g[n]\equiv g(\Omega[n]).
\end{equation}
 Initially, we set each index of $g[n]$ to zero. For each phonon frequency in the dispersion $\omega_{s,\mathbf{q}}$, we assign it a bin value $n$ corresponding to the value that minimizes $|\Omega[n]-\omega_{s,\mathbf{q}}|$, then increment $g[n]$ by 1/N. After iterating over each value of  $\{s,\mathbf{q}\}$, the discrete DOS $g[n]$ corresponds to the density of phonon states per unit volume at each frequency $n\Delta \omega$. Finally, note that the frequencies of phonons with wave vectors that are related by crystal point group operations are identical. Hence, the sum in Equation \ref{DOS} need only be performed over wave vectors in the irreducible Brillouin zone, multiplied by their appropriate weights. A detailed discussion of zone reduction schemes for various Bravais lattices can be found in Appendix 1 of Wallace.\cite{wallace} For visualization purposes (and for comparison with experiments), the phonon DOS is usually broadened with an appropriate function (a Gaussian or Lorentzian, for example). The tetrahedron method is an especially powerful technique for reliably processing DOS data.\cite{lehmann72,yates07} The DOS used as input to a Helmholtz free energy minimization should not be broadened, however. 
  
\subsubsection{\label{sec:si_chooseT_minF} Choose T, minimize F as a function of lattice parameters}
\begin{figure}
    \centering
    \includegraphics[width=\columnwidth]{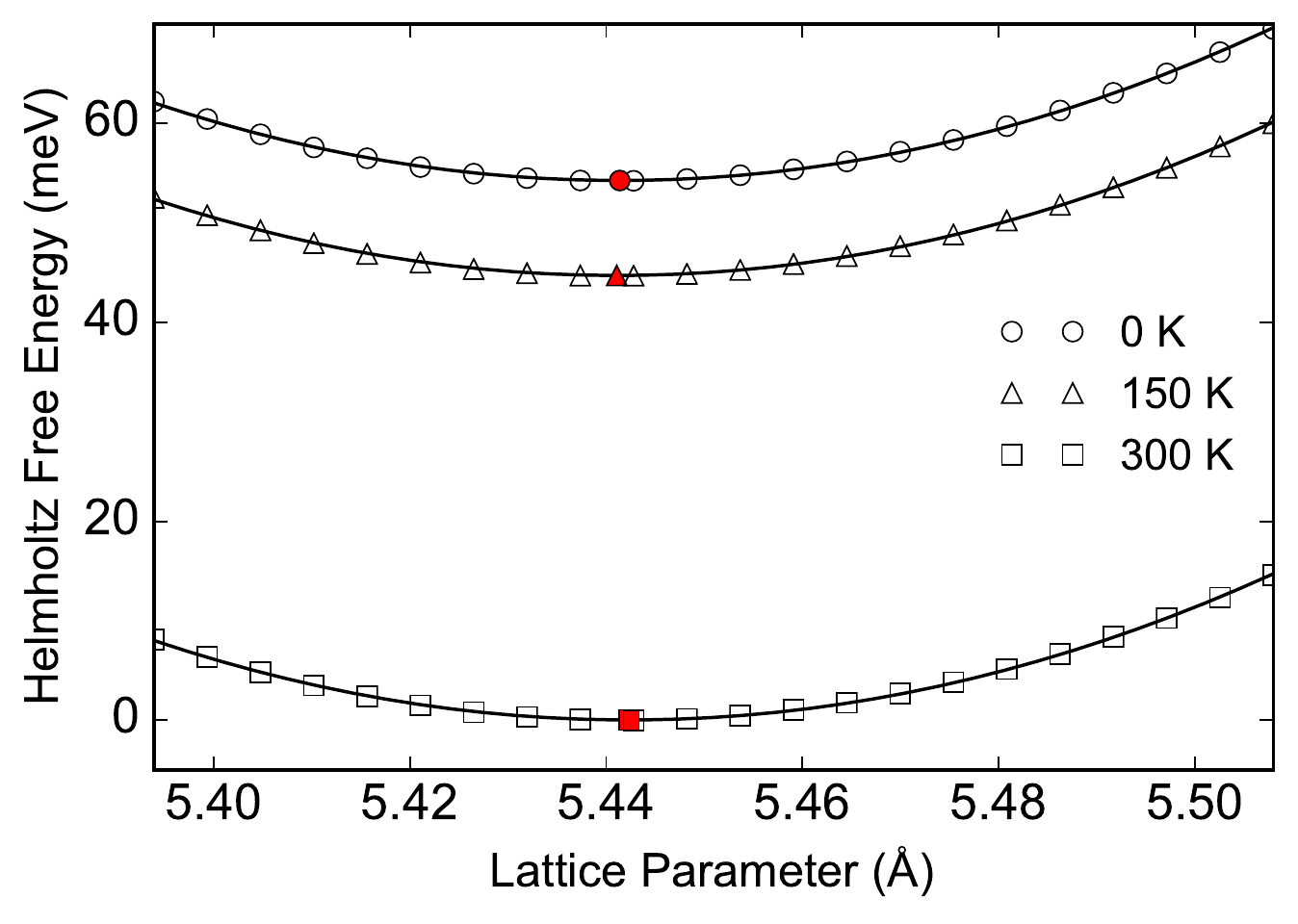}%{19_0617Si_qha_energy_vs_latt_nobar_7.pdf}
    \caption{Helmholtz free energy of silicon from quasiharmonic approximation calculations for three different temperatures from DFT and DFPT using the PBEsol functional. The solid lines indicate second-order polynomial fits and the red markers indicate the equilibrium lattice parameter for each temperature.}
    \label{HelmSi}
\end{figure}

We now have the essential ingredients required for calculating the Helmholtz free energy for a series of unit cells with varying lattice parameters. The total energy from DFT calculations for each set of lattice parameters can be combined with the phonon DOS to calculate Equation \ref{quasiharmDOS} for a given temperature $T_0$. Since silicon is cubic, it has only one lattice parameter, so the generated data will be one-dimensional ($F(a,T_0)$ as a function of $a$). Figure \ref{HelmSi} shows a least-squares polynomial fit to the data. The lattice parameter that minimizes $F(a,T_0)$ at each temperature was found using standard gradient descent methods.

Although the fitting in Figure \ref{HelmSi} may appear simple, finding a good polynomial fit for some systems can be challenging. A useful strategy is to start with the 1 K system (as the log term in Equation \ref{quasiharmDOS} is undefined at $T=0$), perform a polynomial fit, and then use those parameters as initial guesses for the fit of $F(a,T)$ at another temperature $\Delta T$, with $\Delta T$ on the order of 10-20 K. Then, after fitting the system at $\Delta T$, use these new parameters as initial guesses for the system at $2\Delta T$, $3\Delta T$, and so on. In this way, the lattice parameters as a function of $T$ at intervals of $\Delta T$ can be systematically constructed. 

At this point, it is prudent to check whether the chosen range of lattice parameters sufficiently sample the energy surface. If the lattice parameters as a function of temperature are very close or equal to the lattice parameters of the most negatively or most positively strained unit cell, then additional strained unit cells must be added to the analysis until the predicted lattice parameters are well within these lower and upper bounds. 

As an aside to avoid a common source of confusion, note that due to the vibrational zero-point energy introduced by Equation \ref{quasiharmDOS} (the $\frac{\hbar\omega}{2}$ term), the lattice parameters predicted at temperatures close to 0 K using the quasiharmonic approximation are often different than the equilibrium lattice parameters from DFT total energy calculations. This is a correction to the total energy introduced by taking vibrational degrees of freedom into account, and is not an error.

\subsubsection{\label{sec:plot_to_find_alpha} Plot lattice parameters versus T and find $\alpha(T)$}
\begin{figure}
    \centering
    \includegraphics[width=\columnwidth]{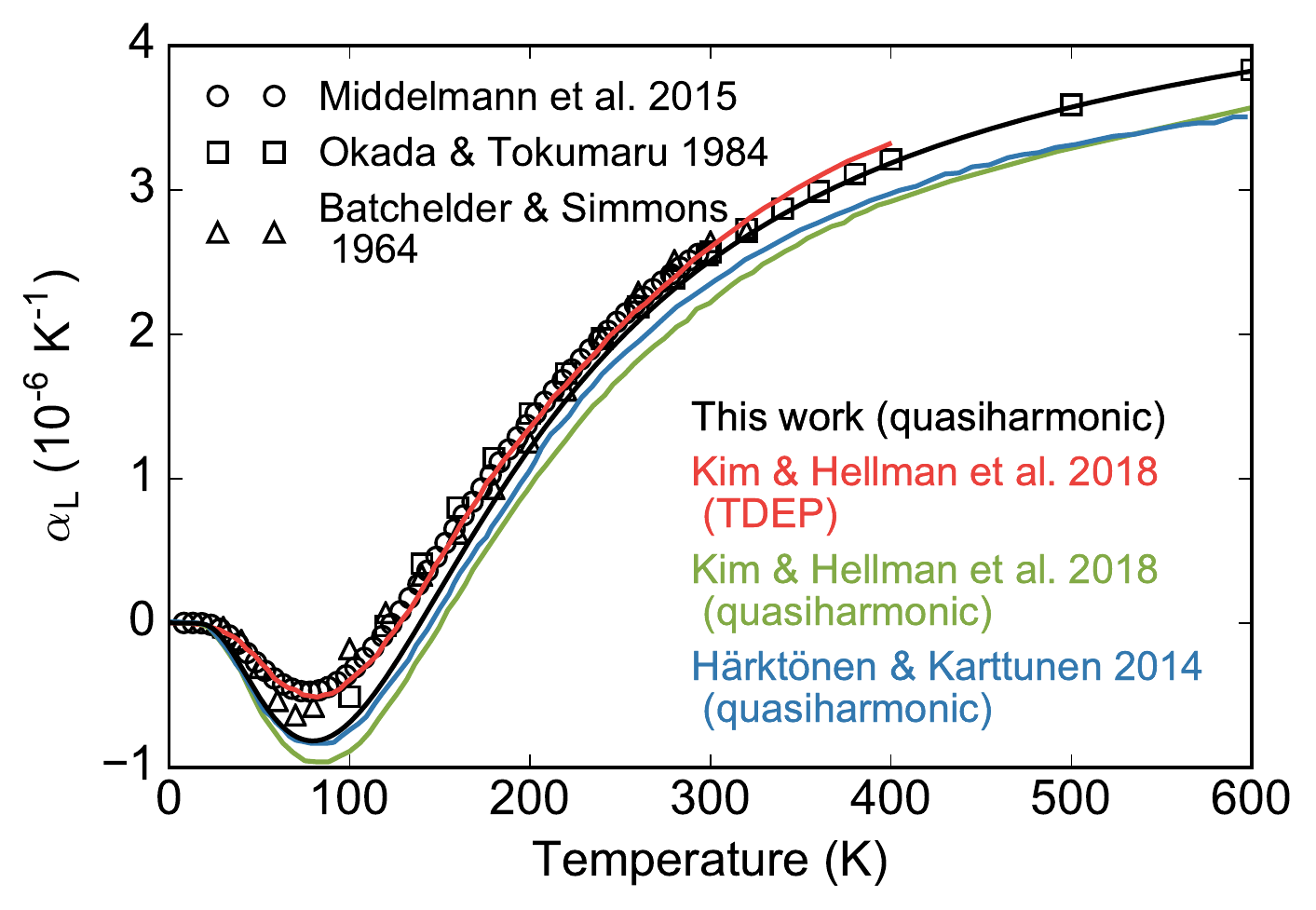}%{19_0708BAs_alpha_latt_vs_temp_numerical_compare.pdf}
    \caption{Linear coefficient of thermal expansion of silicon from first-principles calculations\cite{kim2018nuclear,harkonen2014ab} and experiments (circle,\cite{middelmann2015thermal} square,\cite{okada1984precise} and triangle\cite{batchelder1964lattice}). The black line is calculated using (quasiharmonic) Helmholtz free energy minimization and uses the PBEsol functional, whereas the green\cite{kim2018nuclear} uses the Armiento and Mattsson 2005 (AM05) functional, and the blue\cite{harkonen2014ab} uses the local-density approximation (LDA) functional. The slight disagreement in the three quasiharmonic calculations is due to the functionals. The Temperature Dependent Effective Potential (TDEP) method is discussed further in Section \ref{sec:Limitations}.}
    \label{fig:Si_linear_coeff_thermal_expansion_vs_temp}
\end{figure}

The final step is simply to plot the minimum-energy lattice parameters for each temperature and obtain $\alpha_v(T)$, Figure \ref{fig:Si_linear_coeff_thermal_expansion_vs_temp}. As discussed earlier, the coefficient of volumetric thermal expansion is usually reported in units of inverse temperature, representing volumetric strain per degrees Kelvin, and is proportional to the slope of $V(T)$. In infinitesimal strain theory, the volumetric strain is equivalent to the trace of the strain tensor, $\Delta V = \mathrm{Tr}(\underline{\varepsilon})$. The derivative of the volumetric strain can be found using a simple finite difference at regular intervals of $T$, i.e. for a discrete series of volume as a function of temperature $V[T]$ calculated at regular temperature intervals $\Delta T$,

\begin{equation}
    \alpha_v[T]=\frac{1}{2\Delta T}\frac{V[T+1]-V[T-1]}{V[T]}.
\end{equation}

{}

\subsection{\label{sec:gruneisen} Thermal Expansion from Gr\"uneisen Theory}
We now turn to the Gr\"uneisen framework of thermal expansion, which we can use to interpret results from our free energy calculations for the lattice parameter as a function of $T$ to gain insight into the driving mechanism of thermal expansion.

\subsubsection{\label{sec:si_calcGrun} Select strains and calculate Gr\"uneisen parameters using finite differences}
Both the generalized mode Gr\"uneisen parameter in Equation \ref{eq: grun_aniso} and volumetric mode Gr\"uneisen parameter in Equation \ref{grun_V} require finding the derivative of phonon frequency with respect to a distortion of the unit cell. This can be accomplished using finite differences to find a numerical derivative, which requires three separate phonon dispersion calculations: the unstrained crystal system, and the two systems strained by $\pm \varepsilon$. For each of the strained systems, any internal degrees of freedom must be allowed to relax before performing the phonon calculation (there are no internal degrees of freedom in the case of silicon). 

\begin{figure}
    \centering
    \includegraphics[width = \columnwidth]{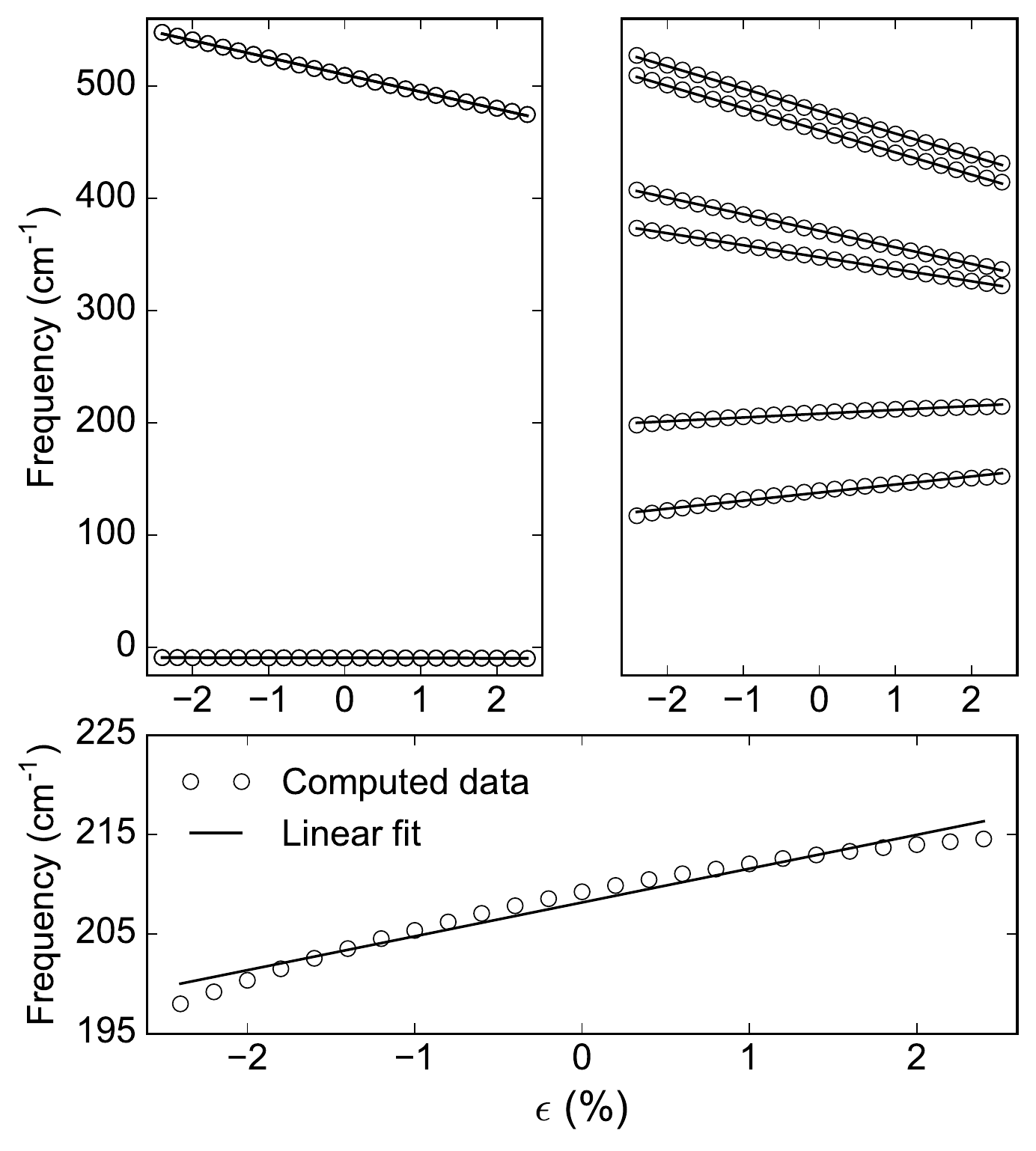}
    %{19_0626_Si_strain_phonons_three_figures_same_y_hide_a2_27.pdf}
    \caption{Phonon frequency response to strain (isotropic volume change) at the $\Gamma$ point (left) and the X point (right) for silicon from DFPT using the PBEsol functional. There are two sets of triply degenerate phonon modes at the $\Gamma$ point. At the lower-symmetry X point, the degeneracies are lifted such that there are six distinct modes. Some of the phonon frequencies exhibit nonlinear behavior when strained beyond $\pm$1\% (bottom).}
    \label{fig:Si_strain_phonons}
\end{figure}

\begin{figure}
    \centering
    \includegraphics[width = \columnwidth]{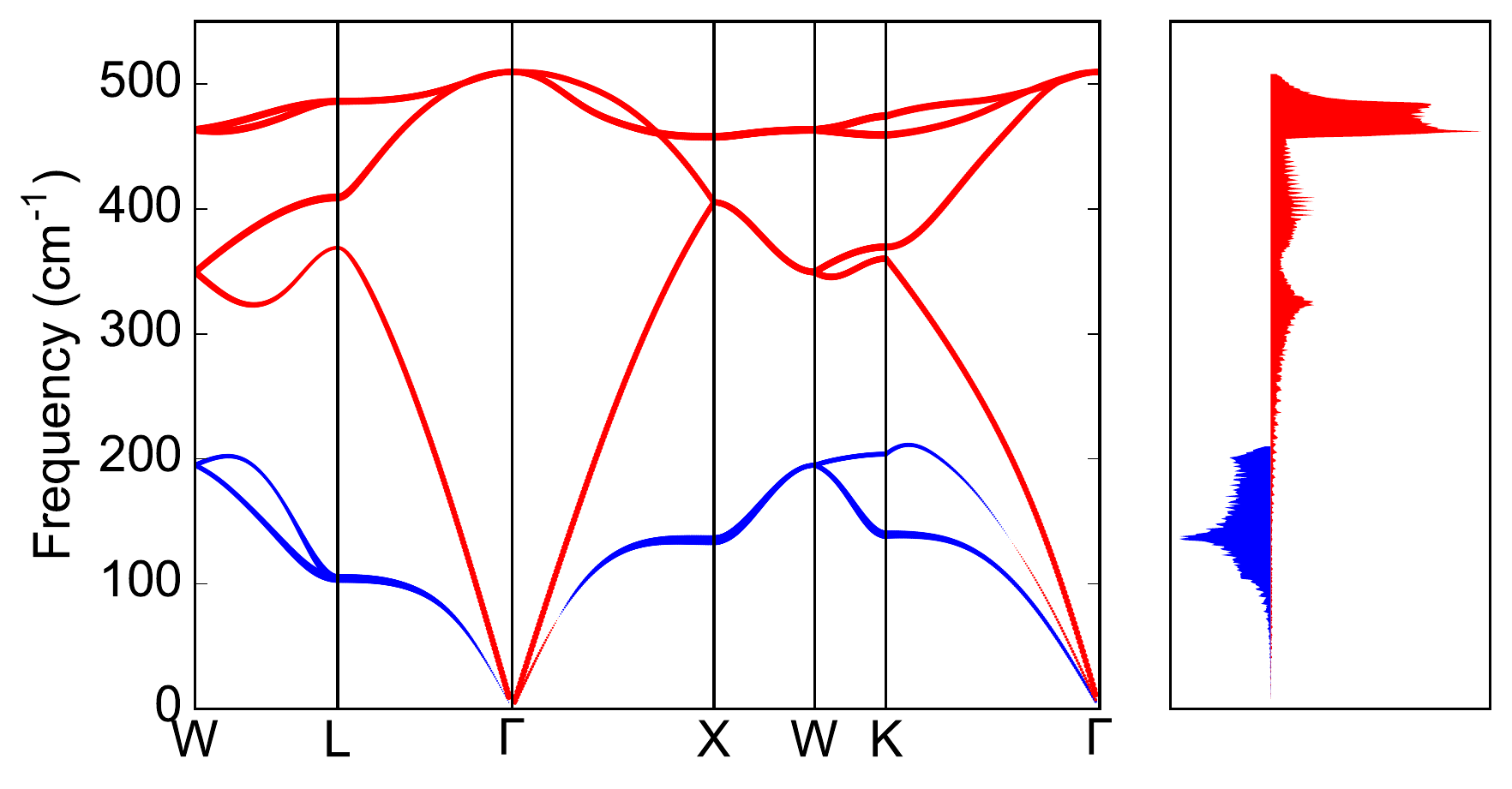}%{19_0807_Si_227_gru_150_dos_strain4_0.pdf}
    \caption{Phonon dispersion curves for silicon from DFPT calculations (using the PBEsol functional) at 0 K with thickness and color of the band proportional to the magnitude and sign of $\gamma^V_{s,\mathbf{q}}$ (red positive, blue negative) calculated at $\epsilon = \pm 0.4\%$ . The sum of $\gamma^V_{s,\mathbf{q}} c_{s,\mathbf{q}}$ across the entire Brillouin zone for each energy level is shown at right, with positive (red) and negative (blue) contributions plotted separately.}
    \label{fig:Si_gru}
\end{figure}

Phonon frequencies typically vary linearly for small strains before anharmonicity contributes to a higher-order response, as shown in Figure \ref{fig:Si_strain_phonons}. However, although we ultimately need to work in the linear response regime, strains that are too small may result in phonon frequency changes that are of the same order as numerical noise. A suitable strain range can be identified by straining the material into a region where the phonon response becomes nonlinear, and then selecting strains that lie comfortably within the linear response range. Ideally, this series of calculations would be performed for all phonons in the system throughout the Brillouin zone but this may not be practical for complex materials or those with large unit cells. The phonon frequency response should at least be checked at a few high-symmetry wave vectors and at least one low-symmetry wave vector. For silicon, we observe a nonlinear response to strain (isotropic volume change) around $\pm$1\% for some modes. Hence, we used strains of $\epsilon= \pm0.4\%$ to calculate the mode Gr\"uneisen parameters. The results can be effectively visualized for simple systems by plotting a phonon dispersion curve using color to indicate the sign of the Gr\"uneisen parameter and line width to indicate its magnitude, as shown in Figure \ref{fig:Si_gru} for silicon. Further details concerning the calculation of Gr\"uneisen parameters are discussed in the next section.

{}

{}

\subsubsection{Calculate heat capacities and bulk modulus}
The expression for the volumetric coefficient of thermal expansion, Equation \ref{alpha_vol}, contains the \emph{bulk} Gr\"uneisen parameter, the specific heat capacity at constant volume, and the bulk modulus. The bulk Gr\"uneisen parameter is essentially a specific heat-weighted sum of the mode Gr\"uneisen parameters across the Brillouin zone. In the case of silicon, a face-centered cubic material, this involves sampling one of 48 equivalent irreducible wedges of the first Brillouin zone, enclosed by the polyhedron $\Gamma$-X-W-K-L-U, as outlined in Wallace.\cite{wallace} The coordinates for the high-symmetry wave vectors are given in the caption of Figure \ref{fig:Si_phonon_disp} and the U point is (2$\pi/a$)(0.25, 1, 0.25). The wave vectors must be assigned their correct weights depending on their location on the corners, edges, surfaces, and interior of the irreducible wedge. Since the Gr\"uneisen parameters are calculated using finite differences at three different volumes, the reciprocal space volume also obviously changes. It is important to ensure that the same number of wave vectors are sampled for the three different volumes. Rather than comparing the change in frequency at each wave vector, we compare phonon frequencies at the same fractional portion of reciprocal space.

The heat capacity at constant volume, $C^V$, is also required to predict the coefficient of thermal expansion under the Gr\"uneisen framework. The form of $C^V$ follows from our assumption of the quasiharmonic approximation for the Helmholtz free energy. Since,\cite{grimvall1999thermophysical} 
\begin{equation}
    C^V\equiv T\biggr(\frac{\partial S}{\partial T}\biggr)_V,
\end{equation}
and since $S$ is determined by the free energy function, $C^V$ is defined through the entropic term of Equation \ref{quasiharm}. The result can be found in many textbooks, \cite{kittel1976introduction} and is given by, 
\begin{multline}
    C^V=\frac{\hbar^2}{k_BT^2}\sum_{\mathbf{q},s}\frac{\omega_{\mathbf{q},s}^2e^{\frac{\hbar \omega_{\mathbf{q},s}}{k_BT}}}{(e^{\frac{\hbar \omega_{\mathbf{q},s}}{k_BT}}-1)^2}\\=k_B\int \mathrm{d}\omega g(\omega)\frac{x^2e^{x}}{(e^x-1)^2},
    \label{capacity}
\end{multline}
where $x={\hbar \omega /k_bT}$, and $g(\omega)$ is the density of phonon states discussed in Section \ref{sec:DOS}, each computed at volume $V$. This is the heat capacity at constant volume for the Einstein model of independent oscillators, each with frequency $\omega$. For non-cubic systems, the heat capacity at constant \emph{configuration}, $C^{\eta}$, is required, which is equivalent to $C^{V}$ in cubic systems. The mode specific heat, or the heat capacity arising from a single phonon mode $\omega_{\{\mathbf{q},s\}}$ is required to calculate the bulk Gr\"uneisen parameter. This is simply the contribution of a single element $\{\mathbf{q},s\}$ to the sum of Equation \ref{capacity}. 

Finally, we initially defined $B_T$ in Equation \ref{bulkMod} as the derivative of pressure with respect to volume at temperature $T$; this is not particularly convenient for first-principles calculations. However, since pressure is defined as, \cite{grimvall1999thermophysical}

\begin{equation}
    P=-\biggr(\frac{\partial F}{\partial V}\biggr)_T,
\end{equation}

we can write the bulk modulus as,

\begin{equation}
    B_T=\frac{1}{V}\biggr(\frac{\partial^2F}{\partial V^2}\biggr)_T.
    \label{bt}
\end{equation}

Recall that our use of the quasiharmonic approximation affects which terms appear in the Helmholtz free energy, and since the Helmholtz free energy appears in Equation \ref{bt}, the consequences of ignoring phonon-phonon coupling extend to the bulk modulus. However, since the bulk modulus must be positive for a stable system, the choice cannot  qualitatively affect the thermal expansion, that is, it cannot change its sign. Additionally, the change in bulk modulus as a function of temperature is typically quite small at low temperatures \cite{born1954dynamical,giustino2014materials} as long as the system is not approaching a structural phase transition. For example, the bulk modulus of silicon decreases by only about 1\% over the range of 77 K to 298 K. \cite{mcskimin1964elastic} Hence, it is often a reasonable approximation to use the 0 K bulk modulus for finite but low temperatures, especially if the Gr\"uneisen framework is being employed as an analysis rather than predictive tool. Assuming that the zero-point correction to the total free energy is small, the Helmholtz free energy $F$ at 0 K can be approximated by the ground state electronic energy, $\Phi_0$ of Equation \ref{quasiharm}:

\begin{equation}
    B_0\approx \frac{1}{V}\biggr(\frac{\partial^2\Phi_0}{\partial V^2}\biggr).
\end{equation}

For a cubic system, this derivative is simple. First, we calculate the total energy for a set of isotropically strained unit cells, centered about the equilibrium structure. This can be done by varying the lattice parameter by a total range of, for example, $\pm 1\%$ over 20 increments of $.1\%$ strain, then calculating the total energy for series of unit cells corresponding to each of these lattice parameters. Then, the bulk modulus can be found by the quadratic term in a polynomial fit of the total energy of these unit cells calculated with DFT as a function of volume (or, alternatively, by a fit to the Birch-Murnaghan equation of state \cite{birch1947finite}, which explicitly includes the derivative of $B_T$ with respect to pressure). The exact range and number of increments can be adjusted until a high-quality polynomial fit is achieved.

\subsubsection{\label{sec:Estimating TE} Calculate $\alpha_v(T)$}
As explained in Section \ref{sec:gruneisen}, the coefficient of volumetric thermal expansion, $\alpha_v(T)$, can be estimated at finite temperature using the Gr\"uneisen parameters, heat capacity, and elastic constants. For a cubic system, the problem reduces to one dimension, and we simply find the appropriate quantities at temperature $T$ at the lattice parameters predicted in Section \ref{sec:TE_from_helm} and evaluate Equation \ref{alpha_vol} to find $\alpha_v(T)$. At a given temperature $T$, there should be good agreement between the $\alpha_v(T)$ calculated under the Gr\"uneisen framework and the $\alpha_v(T)$ calculated using the explicit minimization of the Helmholtz free energy in Section \ref{sec:TE_from_helm}.

\subsection{\label{ComparisonFG} Comparison between Free Energy and Gr\"uneisen frameworks}
At this point, the reader may ask: why bother doing a Gr\"uneisen analysis at all? The Gr\"uneisen framework requires as input the lattice parameters as a function of temperature, which are obtained by minimization of the Helmholtz free energy, as described above. This already provides enough information to calculate $\alpha_v(T)$ directly from the slope. If the Gr\"uneisen framework isn't expected to add qualitative or even quantitative accuracy, then why bother?

We recommend the procedure outlined in Section \ref{sec:TE_from_helm} to calculate thermal expansion coefficients. In addition to being simpler and more robust against numerical errors, the Helmholtz free energy captures the coupling between each phonon and strain to all orders, whereas the Gr\"uneisen framework includes this coupling to first order only. Furthermore, at finite temperatures, the isothermal bulk modulus (or more generally, the elastic constants) must be calculated from derivatives of the Helmholtz free energy, rather than the total energy at 0 K. This means that a full phonon dispersion curve must be calculated \emph{at each strain value}.

However, although minimization of the Helmholtz free energy can predict the temperature dependence of structural parameters with good accuracy (in materials for which the quasiharmonic approximation is justified), the results provide little information as to \emph{why} the lattice parameters shrink or expand with temperature, or how each phonon mode contributes to thermal expansion. The Gr\"uneisen framework can provide these mechanistic insights. Equation \ref{alpha_vol} shows that a positive $\gamma^V_{bulk}$ corresponds to positive thermal expansion. Since $\gamma^V_{bulk}$ is the sum of the mode Gr\"uneisen parameters discussed in Section \ref{sec:si_calcGrun}, the Gr\"uneisen framework allows us to elucidate whether an individual phonon mode is pushing the system towards negative or positive thermal expansion: phonon modes with $\gamma^V_{s,\mathbf{q}}<0$ drive the thermal expansion coefficient more negative, whereas modes with $\gamma^V_{s,\mathbf{q}}>0$ drive the thermal expansion coefficient more positive. In Figure \ref{fig:Si_gru}, which shows the sign of the Gr\"uneisen parameter for each phonon mode, we can conclude that the low-frequency transverse acoustic modes affect thermal expansion in a qualitatively different way that the longitudinal acoustic or optical modes. In this way, we can develop a deeper understanding of the behavior of lattice parameters as a function of temperature -- at low temperatures, the low-frequency phonon modes with mostly $\gamma^V_{s,\mathbf{q}}<0$ are more likely to be occupied than the high-frequency modes with mostly $\gamma^V_{s,\mathbf{q}}>0$. This is consistent with the observed trend of NTE at low temperatures in silicon, which gives way to PTE at higher temperatures. By understanding how each mode influences thermal expansion through the Gr\"uneisen framework, we can make informed statements about the underlying microscopic mechanism of thermal expansion in a given system.

\section{\label{sec:Limitations} Assessing the validity of the Quasiharmonic Approximation}

At the beginning of this Tutorial we briefly mentioned two scenarios in which the quasiharmonic approximation is not valid -- at high temperatures approaching the melting point of a given material, and materials that have a dynamical instability at 0 K. However, there may be many materials that are dynamically stable at low temperatures, but in which the phonon-phonon interactions are also strong. Indeed, phonon-phonon interactions are important in \emph{all} materials -- the thermal conductivity would be infinite without phonon-phonon coupling. But how can we tell when phonon-phonon coupling is too strong to justify the quasiharmonic approximation? The answer depends, to some extent, on what we hope to gain from the thermal expansion study.

Perhaps we wish to use the quasiharmonic approximation and the procedure outlined above to explore the microscopic mechanism underlying the thermal expansion in a given material. If experimental data concerning the structural, elastic and vibrational properties of the material exist, then it will provide an important check on the reliability of our model. If the structural parameters of the material of interest have been measured as a function of temperature, and the quasiharmonic approximation fails to qualitatively reproduce the observed thermal strain, then it obviously cannot be a good tool for understanding thermal expansion in this particular system. However, qualitatively reproducing the observed thermal strain is a necessary but insufficient check -- the model could be getting the right answer for the wrong reason. Hence, the temperature dependence of individual phonon frequencies and elastic constants should also be checked, to make sure that the quasiharmonic approximation is capturing trends in temperature correctly. 

IR and Raman experiments are a common source of data on phonon frequencies, however only phonons that obey certain symmetry selection rules can be measured, and most studies are restricted to the zone center. Nonetheless, these optical techniques provide a relatively simple means to obtain phonon frequency shifts as a function of temperature. However, thermal expansion arises from the contributions of phonons throughout the Brillouin zone, and it may be possible to correctly capture the frequency shifts at the zone center, while describing phonons with non-zero wave vectors poorly. Neutron scattering experiments, which are far more difficult to perform than simple optical experiments, can probe wave vectors away from the zone center and can provide a comprehensive source of data for comparison with theory. If the calculated frequencies and their shifts qualitatively match those measured in a neutron scattering experiment, then the quasiharmonic approximation is likely justified for the system of interest. Inelastic X-ray scattering and diffuse X-ray scattering are additional methods that can be used to probe phonon frequencies across the entire Brillouin zone and act as a good check against computational results, though they are relatively new techniques that require access to an appropriately equipped synchrotron or free-electron laser.\cite{AQRBaronIXS,trigo2013fourier,jiang2016origin}

Instead of delving deeply into microscopics, perhaps we wish to use the procedure described above to quickly estimate the thermal expansion properties of a material for which there is little experimental data -- it may be difficult to synthesize, have been recently discovered, or just overlooked. Or perhaps we would like to use the quasiharmonic approximation to rapidly explore a materials design space, and to highlight systems with potentially interesting thermal properties for synthesis or more rigorous modeling. In these cases, we have only general rules of thumb and experience to guide us. For example, additional caution should be exercised when working with materials that undergo structural phase transitions, since the quasiharmonic approximation may be poor in the vicinity of a phase change. Also, since low thermal conductivity is often associated with high phonon-phonon scattering rates through anharmonic terms not captured by the quasiharmonic approximation, materials with low thermal conductivity may be more challenging for the quasiharmonic approximation (though we emphasize that this is a rough generalization). If any scattering experiments that measure phonon frequencies exist, even if they are limited in scope, one might consider the spectral width of phonons in the material in question. The linewidth of a phonon mode is associated with its mean lifetime -- larger linewidths indicate a shorter lifetime, and thus unusually large linewidths could suggest a material with high levels of phonon-phonon scattering. 

Finally, we pause to offer some remarks on the use of the quasiharmonic approximation to study thermal expansion in silicon specifically, the material we have used as our example throughout this Tutorial. Figure \ref{fig:Si_linear_coeff_thermal_expansion_vs_temp} shows that the quasiharmonic approximation within first-principles DFT can qualitatively, even quantitatively, reproduce the observed change in the volume with temperature. But are the phonon frequency shifts correctly captured? In other words, do we get the right answer for the right reasons?

In a recent insightful and comprehensive study of thermal expansion in silicon,\cite{kim2018nuclear} the authors investigate the thermal behavior using two different theoretical techniques (DFT within the quasiharmonic approximation, and DFT within the Temperature Dependent Effective Potential, TDEP, framework \cite{hellman2011lattice,hellman2014phonon,romero2015thermal}) and neutron scattering. They find that even though the thermal expansion is qualitatively similar between all three methods, a close comparison between them shows that the quasiharmonic approximation predicts that phonon frequencies change with temperature in a manner that is quite different than that seen in the TDEP and neutron scattering data. In particular, the quasiharmonic approximation predicts that the frequencies of longitudinal acoustic modes with negative Gr\"uneisen parameters should \emph{increase} with increasing temperature. Since silicon has a positive coefficient of thermal expansion, this prediction makes sense under the assumptions of the quasiharmonic approximation, where the frequency of a phonon mode with negative $\gamma^V_{s,\mathbf{q}}$ should, by definition, increase when the volume increases. However, both TDEP and neutron scattering experiments show that the frequencies \emph{decrease}, suggesting that phonon-phonon coupling, which is not accounted for in the quasiharmonic approximation, contributes significantly to the dynamics in silicon.

It is important to note that the Gr\"uneisen parameters of the longitudinal acoustic modes really are negative -- their frequencies increase when the volume increases -- and this can be confirmed by experiment.\cite{weinstein1975raman} The quasiharmonic approximation therefore captures the coupling to volume correctly. The issue is that as the temperature increases, these modes are not only coupled to volume in a way that pushes their frequencies up, but they are also coupled to other phonons in a way that pulls their frequencies down. In this case, the phonon-phonon coupling is stronger than the phonon-strain coupling, leading to phonon frequency behavior the opposite of that predicted by the quasiharmonic approximation. The fact that the quasiharmonic approximation still qualitatively predicts the thermal expansion of silicon quite well is attributed in Ref. \onlinecite{kim2018nuclear} to fortuitous error cancellation. Though this may be true in the case of silicon, the agreement between results obtained with experiments and with the quasiharmonic approximation is striking and could merit further investigation.

\section{\label{sec:Aniso} Thermal Expansion in Anisotropic Materials}
The material we have followed through this Tutorial so far, silicon, is cubic so the thermal expansion tensor of Equation \ref{thermTensor} reduces to a scalar. Though many technologically important and physically interesting materials are isotropic, there are many more systems for which thermal expansion is much more complex. In this section, we briefly discuss aspects of studying thermal expansion in a non-cubic material, using the ferroelectric phase of PbTiO$_3$ (space group \# 99, $P4mm$) as an example. See Ref. \onlinecite{ritz2018interplay} for details regarding choice of calculation parameters for the results described below.

Since the ferroelectric phase of PbTiO$_3$ is tetragonal, the thermal expansion tensor is not isotropic, and has the form, 
\begin{equation}
\underline{\alpha}= \begin{bmatrix}\alpha_a & 0 & 0 \\ 0 & \alpha_a & 0 \\ 0 & 0 & \alpha_c
\end{bmatrix},
\label{thermTet}
\end{equation}

where $\alpha_{11}=\alpha_{22}=\alpha_a$ and $\alpha_{33}=\alpha_c$. Though it is still true that $\alpha_v=\mathrm{Tr}(\underline{\alpha})$, the strain per degree along each unique axis can have different magnitudes \emph{and} different signs. The form of the thermal expansion tensor is dictated by crystallographic symmetry. For example, in an orthorhombic system, since the lengths of all three lattice vectors defining the unit cell are unique, all three diagonal entries in Equation \ref{thermTet} would be unique. In a monoclinic system, $\underline{\alpha}$ would be a 6 $\times$ 6 tensor, as the angles between lattice vectors described by the lower right-hand block change with temperature. These added degrees of freedom require modifications to the framework described previously for computing thermal expansion. 

\subsection{\label{sec:tet_qha}Calculate thermal expansion from Helmholtz free energy for anisotropic system}
As mentioned earlier, regardless of the material under investigation, or its symmetry, all lattice parameters and atomic positions must be fully relaxed and converged with respect to calculation parameters. Table \ref{tab:lat_PbTiO3} compares the optimized lattice parameters and atomic coordinates of PbTiO$_3$ between first-principles calculations and experiment. In addition, before applying the quasiharmonic approximation, one must confirm that all phonons of the 0 K system are real, just as for the isotropic case.

\begin{table*}
\caption{\label{tab:lat_PbTiO3}Optimized structural parameters for PbTiO$_3$ in the ferroelectric tetragonal phase from first-principles calculations and experiment. The columns labeled Ti$_z$, O$_z$, and O2,3$_z$ correspond to the $z$ value in fractional coordinates of lattice parameter $c$ for titanium, apical oxygen, and planar oxygen, respectively. The DFT functionals are PBEsol, local-density approximation (LDA), and Wu-Cohen (WC).} 
\centering
%\begin{tabular}{c|c|c|c|c|c|c}
\begin{tabularx}{\textwidth}{c|Y|Y|Y|Y|Y|Y}
  \hline
Method & $a$ [\AA]  & $c$ [\AA]  & $c/a$  & Ti$_z$  & O$_z$  & O2,3$_z$\\
  \hline 
DFT, PBEsol \cite{ritz2018interplay}  & 3.863  & 4.239  & 1.097  & 0.540  & 0.124  & 0.627 \\ 
DFT, WC \cite{ritz2018interplay}  & 3.873  & 4.209  & 1.086  & 0.538  & 0.117  & 0.622\\ 
DFT, LDA \cite{garcia1996first} & 3.862 & 4.071 & 1.046 & 0.524 & 0.082 & 0.589 \\ 
DFT, WC \cite{wu2006more} & 3.890 & 4.193 & 1.078 & 0.532 & 0.108 & 0.611 \\ 
Neutron Diffraction \cite{nelmes1985crystal} & 3.902 & 4.156 & 1.065 & 0.538 & 0.112 & 0.617 \\ \hline
 %\end{tabular}
 \end{tabularx}
\end{table*}

\subsubsection{\label{sec:aniso_range} Select range of lattice parameters for anisotropic system}
Since there is only one lattice parameter degree of freedom in a cubic system, the Helmholtz free energy is minimized at a given temperature along a \emph{line} corresponding to different values of the lattice constant. In tetragonal materials, such as PbTiO$_3$, there are two free lattice parameters, and at each temperature the Helmholtz free energy must be minimized across a \emph{surface}, as shown in Figure \ref{fig:PTO_qha_energy_surface}. For non-cubic materials of primitive tetragonal or orthorhombic symmetry, the required changes in the lattice parameters can be generated by simply changing the lengths of orthogonal lattice vectors. The grid should contain points involving changes of two or more lattice parameters simultaneously, since the elastic coupling between the axes may be strong.\cite{bansal16} Selecting appropriate strains is not so straightforward for crystals of lower symmetry. The number of dimensions of $F$ is equal to the number of independent structural degrees of freedom --- this includes the angle between lattice vectors for rhombohedral, monoclinic, and triclinic systems. Additionally, rather than relaxed at each value of strain, some studies \cite{dangic2018coupling} include internal degrees of freedom explicitly as another dimension by which to minimize Helmholtz free energy. For example, we could perform a thermal expansion study with PbTiO$_3$ in which the $z$ positions of titanium and oxygen are varied as well as the strain, resulting in a 5-dimensional surface to be minimized rather than the 2-dimensional surface shown in Figure \ref{fig:PTO_qha_energy_surface}. This adds much computational effort, and will not be demonstrated in this example, but may prove necessary if quantitative accuracy is desired at higher temperatures, especially near phase changes. For now, we proceed by explicitly minimizing free energy as a function of $a$ and $c$ only. 

The range of lattice parameters must be generated in a systematic and consistent way. We describe a general procedure below that can be applied to the study of materials of any symmetry, and that can also be used for calculating Gr\"uneisen parameters and elastic constants (for further details, see the classic text by Nye\cite{nyebook} and the excellent discussion in Finnis\cite{finnisbook}).

Let $\mathbf{a}$ be a lattice vector of the system and $\mathbf{u}$ a displacement vector that carries $\mathbf{a}$ to $\mathbf{a}^\prime$, such that $\mathbf{a}'=\mathbf{a}+\mathbf{u}$. The Lagrangian strain tensor used throughout this Tutorial is defined through its relationship with $\mathbf{u}$ through

\begin{equation}
    \varepsilon_{ij}\equiv\frac{1}{2}\biggr(\frac{\partial u_i}{\partial x_j}+\frac{\partial u_j}{\partial x_i}\biggr).
\end{equation}
Here, $i$ and $j$ correspond to the indices of a $3 \times 3$ tensor, written in the basis $x_i$. In order to eliminate rotations, $\underline{\varepsilon}$ is constructed to be symmetric, resulting in six unique $\varepsilon_{ij}$ that together define a state of homogeneous strain in the system.  When applied to $\mathbf{a}$, $\underline{\varepsilon}$ provides $\mathbf{u}$:

\begin{equation}
\mathbf{u}= \underline{\varepsilon}\mathbf{a}=\begin{bmatrix}\varepsilon_{11} & \varepsilon_{12} & \varepsilon_{13} \\ \varepsilon_{21} & \varepsilon_{22} & \varepsilon_{23} \\ \varepsilon_{31} & \varepsilon_{32} & \varepsilon_{33}
\end{bmatrix}\mathbf{a}.
\end{equation}

The lattice vectors that describe the unit cell of the strained system are then given by,
\begin{equation}
\mathbf{a}'= (\underline{1}+\underline{\varepsilon})\mathbf{a},
\label{applyStrain}
\end{equation}

where $\underline{1}$ is the 3$\times$3 identity matrix. In order to explore a range of strains of the same shape but with different magnitudes, we define a strain tensor for our system such that, 
\begin{equation}
\underline{\varepsilon}(\mu)=\mu \underline{\tau}.
\label{strainmuT}
\end{equation}
Here, $\mu$ is a small unitless scalar parameter that specifies the size of the strain (and can take positive \emph{or} negative values), and $\underline{\tau}$, defined below for a homogeneous strain, specifies the shape:
\begin{equation}
\underline{\tau}= \begin{bmatrix}1 & 0 & 0 \\ 0 & 1 & 0 \\ 0 & 0 & 1
\end{bmatrix}.
\label{diagT}
\end{equation}

By choosing a range of $\mu$ values, one can then generate a range of strains that each correspond to a different unit cell. For example, in the case of PbTiO$_3$, we can define $\underline{\tau}^a$ and $\underline{\tau}^c$, corresponding to strains along the $a$ and $c$ axes, respectively: 
\begin{equation}
\underline{\tau}^a= \begin{bmatrix}1 & 0 & 0 \\ 0 & 1 & 0 \\ 0 & 0 & 0
\end{bmatrix},
\label{Ta}
\end{equation}
\begin{equation}
\underline{\tau}^c= \begin{bmatrix}0 & 0 & 0 \\ 0 & 0 & 0 \\ 0 & 0 & 1
\end{bmatrix}.
\label{Tc}
\end{equation}

Then, to generate a series of unit cells on which to perform the quasiharmonic approximation, we can generate a series of strains such that,
\begin{equation}
\underline{\varepsilon}(\mu_a,\mu_c)=\mu_a \underline{\tau}^a+\mu_c \underline{\tau}^c,
\label{strainmuT}
\end{equation}
using a grid of $\mu_a$ and $\mu_c$ values, and apply $\underline{\varepsilon}(\mu_a,\mu_c)$ to the unit cell using Equation \ref{applyStrain}.

Finally, before using these strained unit cells to evaluate phonon frequencies, unless the internal degrees of freedom are being treated as independent parameters with which to optimize the Helmholtz free energy, the positions of the atoms must be allowed to relax to equilibrium (the forces must be minimized).

\subsubsection{\label{sec:aniso_range} Calculate phonon dispersion curve, DOS, and Helmholtz free energy for all lattice parameters, then calculate thermal expansion tensor}
\begin{figure}
    \centering
    \includegraphics[width = \columnwidth]{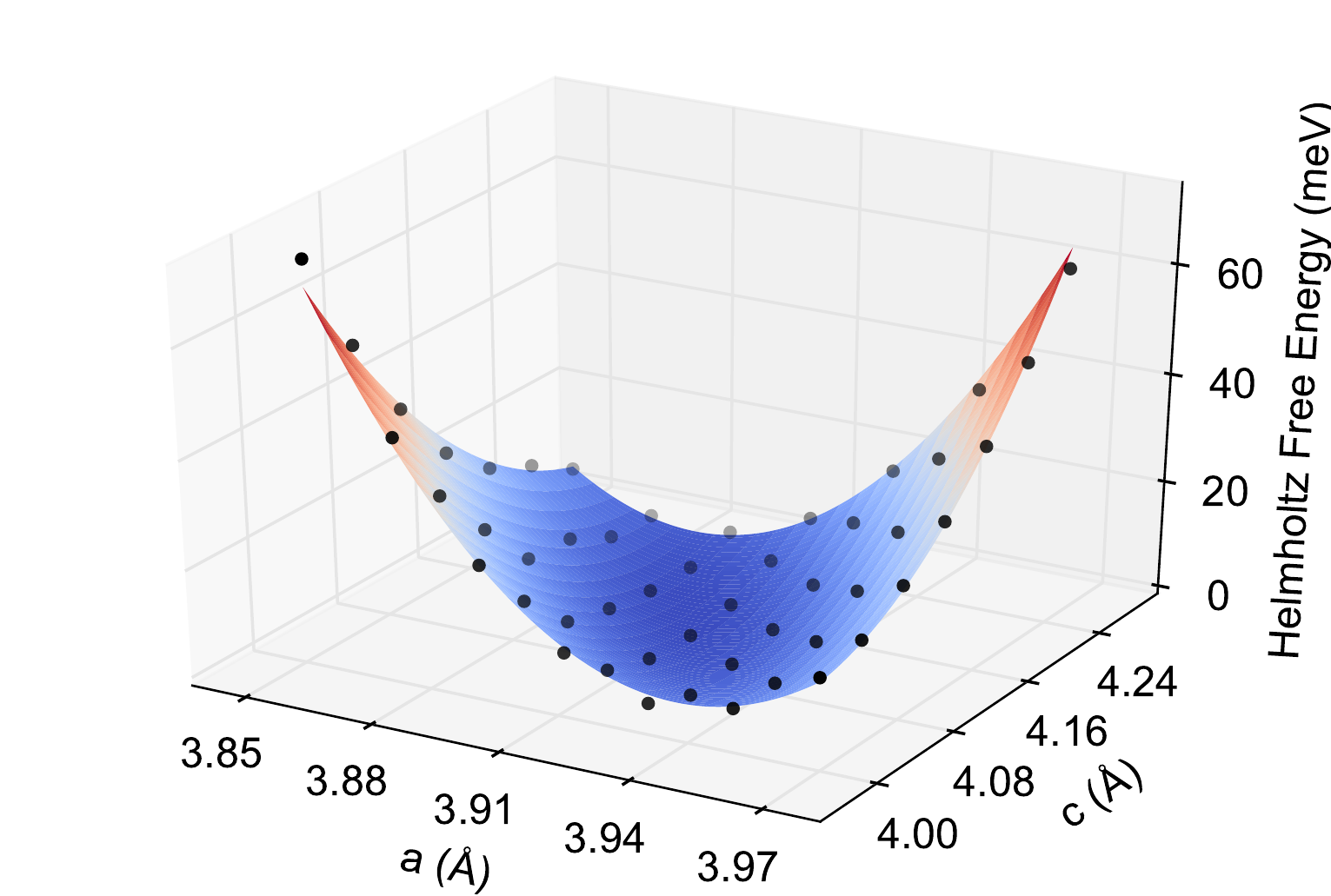}%{19_0626_PTO_qha_F_surface_meV5.pdf}
    \caption{Helmholtz free energy at 300 K with respect to strain for PbTiO$_3$. The energy surface is fitted to the calculated black points. Calculations use the Wu-Cohen functional.}
    \label{fig:PTO_qha_energy_surface}
\end{figure}

\begin{figure}
    \centering
    \includegraphics[width=\columnwidth]{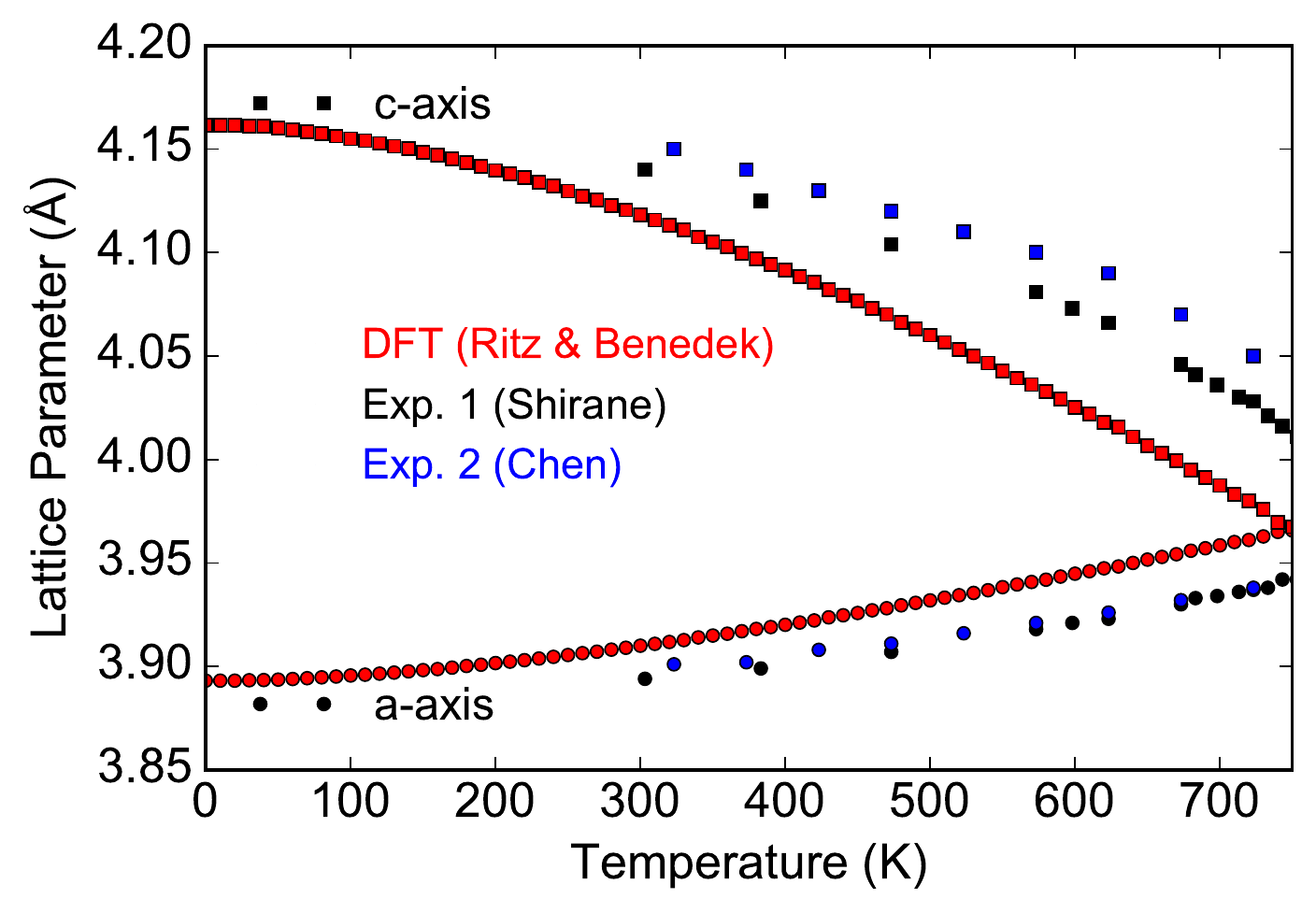}%{19_0626_Ethan_fig1_latt_v_temp8.pdf}
    \caption{Lattice parameters ($a$ circles, $c$ squares) as a function of temperature from first-principles Helmholtz free energy minimization calculations using the Wu-Cohen functional\cite{ritz2018interplay} and from experiment (black\cite{shirane1951phase} and blue\cite{chen2005structure} icons). Reproduced with permission from Phys. Rev. Lett. 121 255901 (2018). Copyright 2018 American Physical Society.}
    \label{fig:PTO_latt_vs_temp}
\end{figure}

The phonon dispersion curve, phonon density of states, and Helmholtz free energy at each point on the grid of strained unit cells can be calculated using the methods described in Sections \ref{sec:phonondisp_qha}, \ref{sec:DOS}, and \ref{sec:si_chooseT_minF}. Figure \ref{fig:PTO_qha_energy_surface} illustrates both the grid of lattice parameters, as well as the free energy calculated at each configuration. As in Figure \ref{HelmSi}, each point in Figure \ref{fig:PTO_qha_energy_surface} requires a full phonon dispersion calculation at the different values of the $a$ and $c$ lattice parameters using the calculated phonon density of states and Equation \ref{quasiharmDOS}. One practical difference between this procedure in the isotropic and anisotropic cases is that it is much more challenging to find a good polynomial fit for a multidimensional surface than for a line. Many different fitting and optimization algorithms are available, however regardless of the choice of algorithm, a good practice (as mentioned in Section \ref{sec:si_chooseT_minF}) is to use the fitting parameters from the lower temperature surface as initial guesses to fit the higher temperature surface. The lattice parameters corresponding to the minimized Helmholtz free energy in PbTiO$_3$ from 0 to 750 K are plotted in Figure \ref{fig:PTO_latt_vs_temp}. The calculated value of $\alpha_v$ from the DFT data using the derivative of these lattice parameters as in Section \ref{sec:plot_to_find_alpha} is -2.29 $\times$ 10$^{-5}$ K$^{-1}$ between 500 and 700 K, which compares favorably with  $\alpha_v$=-1.8 $\times$ 10$^{-5}$ and -1.99 $\times$ 10$^{-5}$ K$^{-1}$ from Refs. \onlinecite{shirane1951phase} and \onlinecite{chen2005structure}, respectively. The increasing disagreement between theory and experiment as temperature increases could possibly be reduced by including the internal atomic degrees of freedom in the surface used to minimize the Helmholtz free energy, as in Ref. \onlinecite{dangic2018coupling}.

{}

{}

\subsection{\label{sec:tet_gru_theory} Thermal Expansion from Gr\"uneisen Theory for Anisotropic systems}
The additional degrees of freedom introduced by the anisotropy of the unit cell significantly complicate the Gr\"uneisen framework. Most importantly, the generalized tensor form of the mode Gr\"uneisen parameter must be used (Equation \ref{eq: grun_aniso}), rather than the simple volume derivative (Equation \ref{grun_V}). We reiterate that since there is no unique deformation corresponding to a given volume change in anisotropic systems, Equation \ref{grun_V} is poorly defined. Moreover, the tensor form allows for the exploration of the behavior of each structural and vibrational degree of freedom in far more detail than the simple scalar form, giving us an understanding of how the anharmonicity of each phonon mode contributes to the thermal expansion of each lattice parameter individually. In addition, it is no longer sufficient to consider the bulk modulus. The full elastic compliance tensor must be calculated to apply the Gr\"uneisen framework to anisotropic systems.

\subsubsection{\label{sec:tet_gru_param} Select strains and calculate Gr\"uneisen parameters for anisotropic system}
\begin{figure}
    \centering
    \includegraphics[width=\columnwidth]{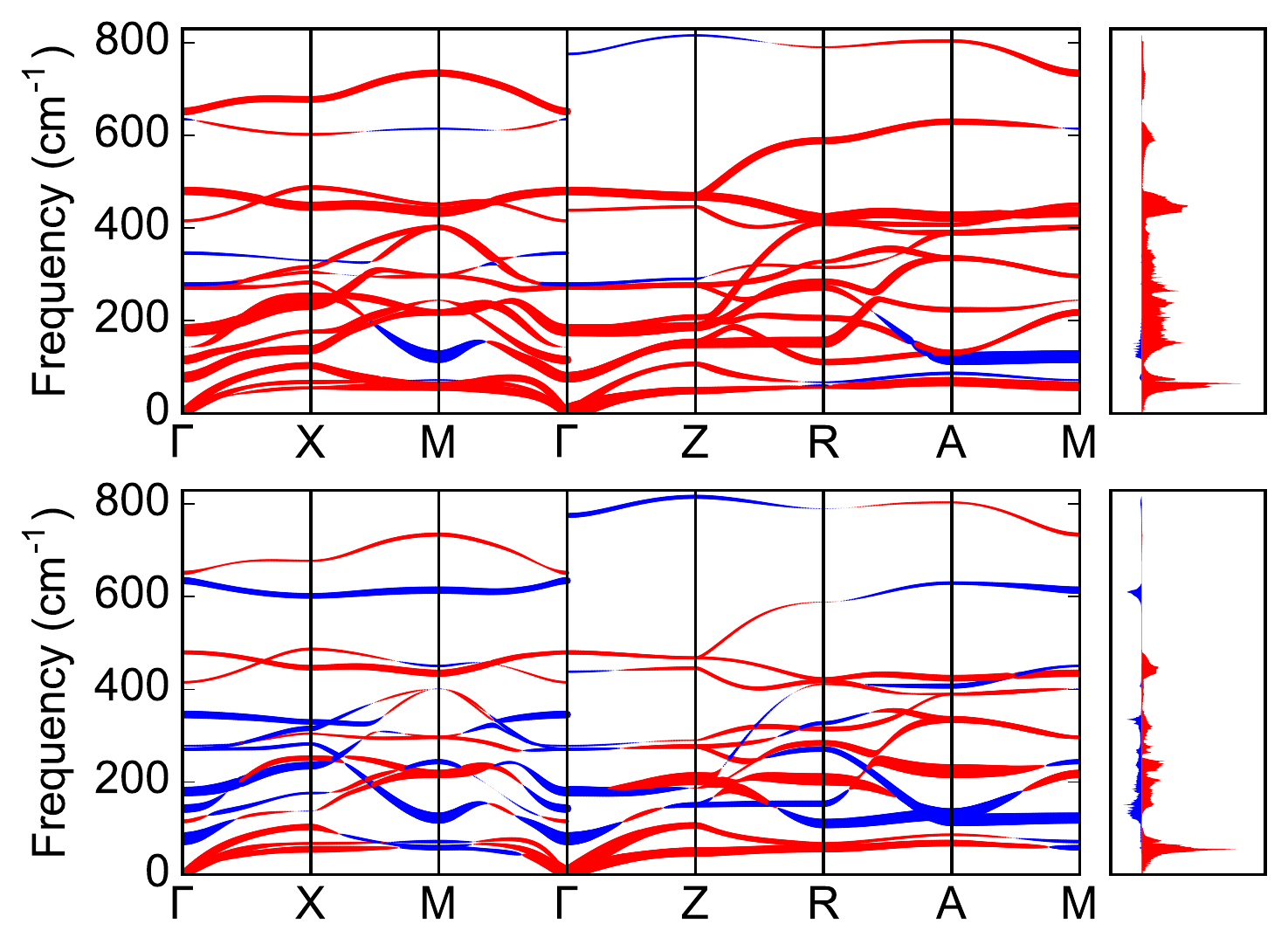}%{19_0626_Ethan_fig2_gru_1.pdf}
    \caption{(Top) Phonon dispersion curves from first-principles DFT (using the Wu-Cohen functional) for PbTiO$_3$ at 300 K with thickness and color of the band proportional to the magnitude and sign of $\gamma^a_{s,\mathbf{q}}$ (red positive, blue negative). To the right is the sum of $\gamma^a_{s,\mathbf{q}} c_{s,\mathbf{q}}$ across the entire Brillouin zone for each energy level; positive (red) and negative (blue) contributions are plotted separately. (Bottom) Same, corresponding to $\gamma^c_{s,\mathbf{q}}$. Reproduced with permission from Phys. Rev. Lett. 121 255901 (2018). Copyright 2018 American Physical Society.}
    \label{fig:PTO_gru_phonon_disp_a_c}
\end{figure}

The procedure here is essentially the same as that described in Section \ref{sec:si_calcGrun}, however additional calculations are required to account for the extra structural degrees of freedom. Since PbTiO$_3$ has two independent lattice constants, it has two unique bulk Gr\"uneisen parameters -- one describing the behavior of phonons with respect to strains along the $a$ axis, and one with respect to the $c$ axis. In general, a phonon will have one Gr\"uneisen parameter corresponding to each unique degree of freedom in the thermal expansion tensor of Equation \ref{thermTet}, though it is also possible to expand the Gr\"uneisen framework to include additional internal degrees of freedom in systems where these would provide particularly important insight.\cite{dangic2018coupling} For PbTiO$_3$, strains of $\epsilon = \pm0.2\%$ were used to calculate the generalized mode Gr\"uneisen parameters $\gamma^a_{s,\mathbf{q}}$ and $\gamma^c_{s,\mathbf{q}}$  in the phonon dispersion curves of Figure \ref{fig:PTO_gru_phonon_disp_a_c}, and the respective bulk Gr\"uneisen parameters along those crystallographic directions. As mentioned before, any internal degrees of freedom must be allowed to relax before performing the phonon calculations for the strained system (in the case of tetragonal PbTiO3, with the position of the Pb nucleus taken as the unit cell origin, the Wyckoff sites of each other nucleus have free parameters along $z$). The components of the bulk Gr\"uneisen parameter for a tetragonal system are the sum of the mode Gr\"uneisen parameters across the Brillouin zone, weighted by $c_{s,\mathbf{q}}$:
\begin{equation}
    \label{eq:tet_bulk_gru}
    \gamma_{bulk}^{ij} = \frac{\sum_{s,\mathbf{q}}\gamma^{ij}_{s,\mathbf{q}}c_{s,\mathbf{q}}}{\sum_{s,\mathbf{q}}c_{s,\mathbf{q}}}.
\end{equation}
For the purposes of evaluating Equation \ref{eq:tet_bulk_gru}, the first Brillouin zone of tetragonal PbTiO$_3$ can be divided into 16 equivalent irreducible wedges enclosed by the triangular prism $\Gamma$-X-M-A-R-Z, with symmetry points defined as $\Gamma$ = 2$\pi$(0,0,0), X = 2$\pi$(0, 1/2$a$, 0), M = 2$\pi$(1/2$a$, 1/2$a$, 0), A = 2$\pi$(1/2$a$, 1/2$a$, 1/2$c$), R = 2$\pi$(0, 1/2$a$, 1/2$c$), Z = 2$\pi$(0, 0, 1/2$c$). The resulting bulk Gr\"uneisen parameters are $\gamma_1 = \gamma^a_{bulk} = 1.42$ and $\gamma_3 = \gamma^c_{bulk} = 0.40$ for a structure with lattice parameters at 300 K, as predicted by the quasiharmonic approximation. 

\subsubsection{\label{sec:tet_gru_compliance}Calculate heat capacity and elastic constants for anisotropic system}
In addition to the bulk Gr\"uneisen parameters, calculation of the thermal expansion coefficient within the Gr\"uneisen framework requires the relevant components of the full compliance tensor -- the bulk modulus is no longer sufficient to characterize anisotropic thermal strain. It can be shown that, for any crystal system, \cite{wallace}

\begin{equation}
\alpha_{ij}=\frac{C_{\eta}}{V}\sum_{kl}S^T_{ijkl}\gamma_{bulk}^{kl},
\end{equation}
where $S^T_{ijkl}$ are the elements of the rank four elastic compliance tensor (the heat capacity at constant configuration, $C_\eta$, is calculated the same way as described earlier). For compactness, $S$ is often expressed as a second order tensor using Voigt notation. This notation can also be used to rewrite the strain $\underline{\varepsilon}$ and Gr\"uneisen tensor $\gamma_{bulk}^{kl}$. Using this notation,

\begin{equation}
\alpha_{i}=\frac{C_{\eta}}{V}\sum_{j}S^T_{ij}\gamma^{j}.
\label{thermInVoit}
\end{equation}

Here $\alpha_i$ and $\gamma^j$ are the elements of 6 $\times$ 1 vectors and $S^T_{ij}$ are the elements of the 6 $\times$ 6 elastic compliance tensor. Note that since the compliance tensor is the inverse of the elastic stiffness tensor $\underline{C}$, one need only calculate $\underline{C}$ for the system of interest and then invert it to obtain $\underline{S}$. 

In order to calculate $\underline{C}$, we first define the strain energy density of a single crystal with volume $V$ by,

\begin{equation}
\frac{E_{strain}}{V}=\frac{1}{2}\sum_{ij}C_{ij}\varepsilon_i\varepsilon_j,
\label{strainEn}
\end{equation}
where we have again used Voigt notation to denote the indices of the strain vector, $\varepsilon_j$. By expanding the sum in Equation \ref{strainEn} and taking into account the symmetry of $\underline{C}$, we obtain a polynomial parameterized by the indices of $\underline{C}$ and strain. For example, as the form of $\underline{C}$ for a tetragonal system is given by,

\begin{equation}
\underline{C}= \begin{bmatrix}C_{11} & C_{12} & C_{13} & 0 & 0 & 0 \\                   C_{12} & C_{11} & C_{13} & 0 & 0 & 0 \\                   C_{13} & C_{13} & C_{33} & 0 & 0 & 0 \\
                  0 &  0 & 0 & C_{44} & 0 & 0 \\
                  0 &  0 & 0 & 0 & C_{44} & 0 \\
                  0 &  0 & 0 & 0 & 0 & C_{66} \\
\end{bmatrix},
\end{equation}
the expansion of Equation \ref{strainEn} takes the form,
\begin{multline}
\frac{E_{strain}}{V}=\frac{1}{2}\biggr(C_{11}(\varepsilon_1^2+\varepsilon_2^2)+C_{33}\varepsilon_3^2+...\\C_{66}\varepsilon_6^2+C_{44}(\varepsilon_4^2+\varepsilon_5^2)\biggr)+C_{12}\varepsilon_1\varepsilon_2+C_{13}(\varepsilon_1\varepsilon_3+\varepsilon_2\varepsilon_3).
\label{strainE}
\end{multline}

We then apply a series of specific strains to isolate each $C_{ij}$. Under the framework of applying strain introduced in Section \ref{sec:aniso_range}, a matrix $\underline{\tau}$ of the form 

\begin{equation}
\underline{\tau}= \begin{bmatrix}0 & 0 & 0 \\ 0 & 0 & 0 \\ 0 & 0 & 1
\end{bmatrix},
\end{equation}
\noindent
will, by Equation \ref{strainmuT}, generate a series of strains that look like,
\begin{equation}
\underline{\varepsilon}= \begin{bmatrix}0 & 0 & 0 \\ 0 & 0 & 0 \\ 0 & 0 & \mu
\end{bmatrix} \implies \varepsilon_i= 
\begin{cases}
    \ \mu,& \text{if } i= 3\\
    0,              & \text{otherwise}
\end{cases}.
\end{equation}

Substituting this into Equation \ref{strainE}, we find
\begin{equation}
\frac{E_{strain}}{V}=\frac{1}{2}C_{33}\mu^2.
\end{equation}
By approximating $E_{strain}$ as the free energy under the quasiharmonic approximation of the strained system for a range of $\mu$ values, the elastic constant C$_{33}$ can be found from the quadratic component of a polynomial fit to the energy as a function of $\mu$. Suitable choices of $\underline{\tau}$ can be used to construct a system of equations that solves for the other elastic constants, and a similar procedure will work to find the full elasticity tensor for systems of any symmetry (see Nye\cite{nyebook} for further details). Alternatively, density functional perturbation theory can be used to calculate elastic constants \cite{baroni2001phonons}, though computational expense scales rapidly with unit cell size.

%Materials with tetragonal symmetry have six unique elastic constants and these can be calculated in the manner described in Section \ref{latRange}, that is, by calculating the Helmholtz free energy with respect to a series of strains. 

\begin{figure}
    \centering
    \includegraphics[width=\columnwidth]{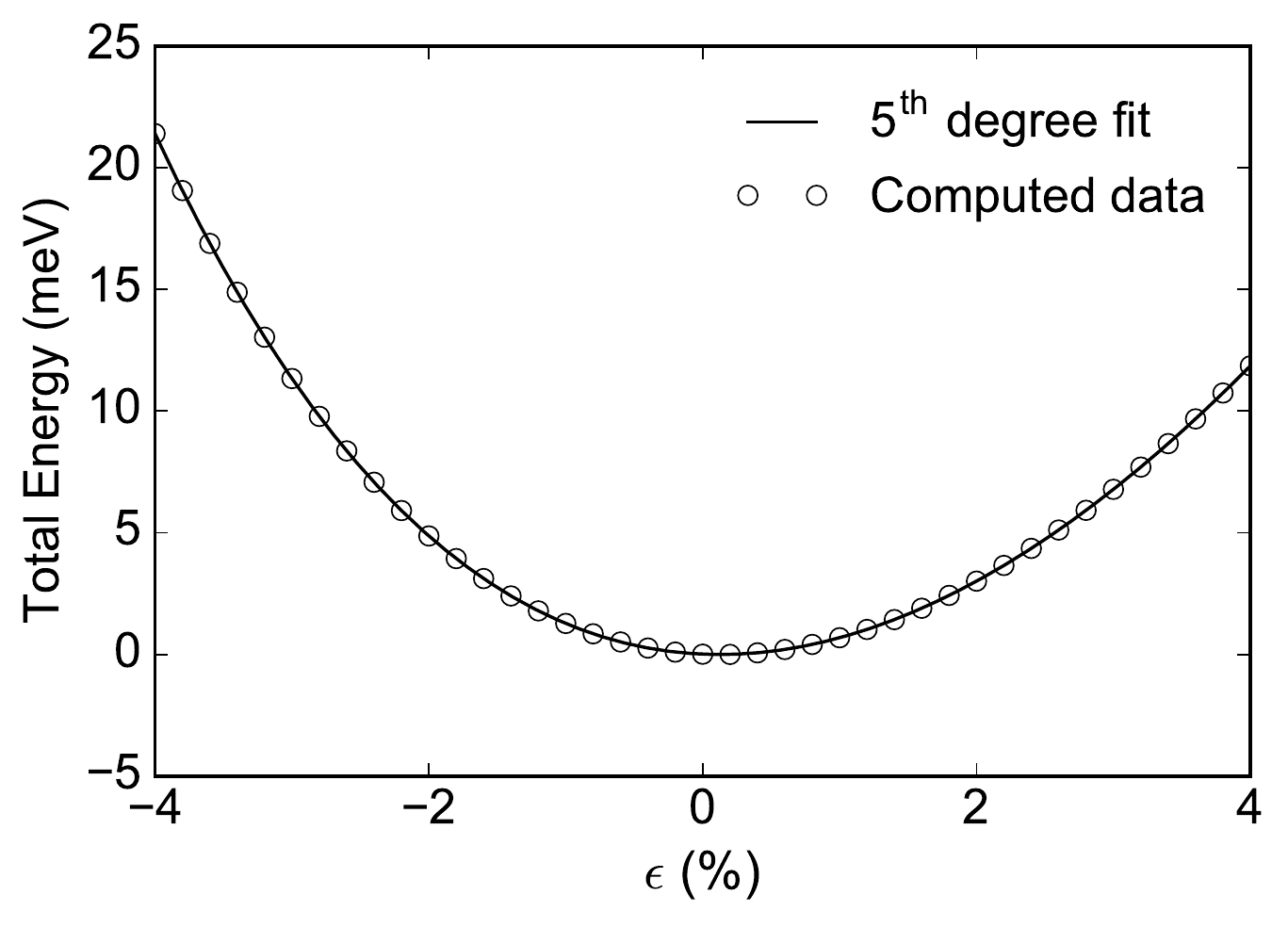}
    %{19_0626_Ritz_PTO_elastic_constant_strain1.pdf}
    \caption{Fifth-degree polynomial fit to energy of PbTiO$_3$ with respect to strain for the $C_{33}$ elastic stiffness constant.}
    \label{fig:PTO_elastic_const}
\end{figure}

In practice, a full quasiharmonic calculation for a single elastic constant requires the calculation of Helmholtz free energy at 10-20 magnitudes of strain, each of which requires a phonon calculation. For the six elastic constants of PbTiO$_3$, this would result in over 100 phonon calculations -- a sizeable computational expense. As discussed earlier, at 0 K the Helmholtz free energy can often be well approximated as the total electronic energy ($\Phi_0$ in Equation \ref{quasiharm}), and those 0 K elastic constants taken to approximate the elastic constants of the material at low temperatures for the purposes of qualitative analysis. Of course, for \emph{quantitative} accuracy, the vibrational contributions to free energy must be taken into account as well.  Figure \ref{fig:PTO_elastic_const} shows a fit to the energy of PbTiO$_3$ with respect to strain for the $C_{33}$ elastic constant from total energy using first-principles calculations at 0 K. The other elastic constants can be obtained in the same way and then inverted to generate the required components of the elastic compliance tensor,
$S_{11} = 7.44$, $S_{12} = 0.49$, $S_{13} = -11.94$, $S_{33} = 55.69$, in units of $10^{-3}$ GPa$^{-1}$.

\subsubsection{\label{sec:tet_alpha} Calculate thermal expansion tensor for anisotropic systems}
 
 Though the relation in Equation \ref{thermInVoit} involves six degrees of freedom, some degrees of freedom may be discarded due to symmetry. For example, in tetragonal systems, $\alpha_4=\alpha_5=\alpha_6=0$, and the only non-zero block of the system of equations described by Equation \ref{thermInVoit} is

\begin{equation}
\begin{bmatrix}\alpha_1\\ \alpha_1 \\ \alpha_3
\end{bmatrix}= \frac{C_{\eta}}{V}\begin{bmatrix}S_{11} & S_{12} & S_{13} \\ S_{12} & S_{11} & S_{13} \\ S_{13} & S_{13} & S_{33}
\end{bmatrix}
\begin{bmatrix}\gamma_1\\ \gamma_1 \\ \gamma_3
\end{bmatrix},
\end{equation}
meaning that 

\begin{equation}
\alpha_1 = \frac{C_{\eta}}{V} \biggr((S_{11}+S_{12})\gamma_1 + S_{13}\gamma_3 \biggr),
\end{equation}
\begin{equation}
\alpha_3 = \frac{C_{\eta}}{V} \biggr( 2S_{13}\gamma_1 + S_{33}\gamma_3 \biggr),
\end{equation}
\begin{equation}
\alpha_v = \frac{C_{\eta}}{V} \biggr(2 (S_{11}+S_{12}+S_{13})\gamma_1 + (2S_{13}+S_{33})\gamma_3 \biggr).
\label{alphavTet}
\end{equation}

Comparing Equations \ref{alphavTet} and \ref{alpha_vol}, it is clear that even adding just one more free lattice parameter can add much more complexity to the thermal expansion behavior. In the isotropic case, the Gr\"uneisen parameter determines the sign of $\alpha_v$, whereas in the tetragonal case, there are seven different quantities ($S_{11}, S_{12}, S_{13}, S_{33}, \gamma_1, \gamma_3$) that can affect the sign of $\alpha_v$. Notably, as explained in Ref. \onlinecite{ritz2018interplay}, $\alpha_v < 0$ when 
\begin{equation}
2 (S_{11}+S_{12}+S_{13})\gamma_1 + (2S_{13}+S_{33})\gamma_3 < 0.
\label{eq:alphav_NTE_cond}
\end{equation}
This equation shows that negative Gr\"uneisen parameters are \emph{not} required for negative thermal expansion in a non-cubic material, and is useful for a qualitative understanding of what drives the sign of volumetric thermal expansion in a tetragonal system. For example, in PbTiO$_3$, the $(2S_{13}+S_{33})\gamma_3$ term in Equation \ref{eq:alphav_NTE_cond} is negative because $S_{13}$ is large and negative, highlighting the importance of this elastic constant in the observed negative thermal expansion behavior, as discussed in Ref. \citenum{ritz2018interplay}. For \emph{quantitatively} accurate calculations of $\alpha_v$ using Equation \ref{eq:alphav_NTE_cond}, $\underline{S}$ must be calculated at finite temperatures, that is, the vibrational contribution to the free energy must be taken into account. However, this would first require a calculation of the lattice parameters as a function of temperature, at which point one can already determine $\alpha_v$. Hence, Equation \ref{eq:alphav_NTE_cond} is most useful as a qualitative model to help interpret the results of quasiharmonic approximation calculations, rather than as a quantitative method for calculation of $\alpha_v$.

\section{\label{sec:summary_outlook}Summary and Outlook}
In this Tutorial, we have described computational procedures for predicting the thermal expansion of insulating materials using first-principles DFT calculations. Two different frameworks were presented, the quasiharmonic approximation and Gr\"uneisen theory, and the advantages and disadvantages of each were discussed.

Perhaps the most challenging aspect of calculating the thermal expansion behavior of materials is understanding when the neglect of phonon-phonon coupling is a justified approximation. Many different techniques have been developed for going beyond the quasiharmonic approximation, however much less effort has been devoted to elucidating the factors that may indicate that use of the quasiharmonic approximation is ill-advised (aside from the obvious cases, mentioned at the beginning of the Tutorial). This is a problem because the quasiharmonic approximation can fail in subtle and complex ways, and because there is currently no real way to check whether a given material may be problematic, without performing a series of computationally expensive phonon calculations. Further work is needed in this area.

Another area requiring further investigation is the study of thermal expansion and the use of the quasiharmonic approximation in materials in which weak interactions contribute significantly to bonding, \textit{e.g.} dispersion-bound molecular crystals.\cite{ko18} Highly anharmonic bonding can lead to significant thermal expansivities in these systems, much greater than those typically observed in inorganic solids. An additional challenge is that dispersion interactions are poorly described by the standard semilocal functionals that are usually used to calculate thermal properties from first principles. Although various correction schemes have been developed, the most appropriate technique for the system of interest is not always obvious.

Finally, a particularly interesting class of materials in which to explore fundamental mechanisms of thermal expansion is metal-organic frameworks.\cite{goodwin08,goodwin08b,goodwin09,cairns12} These systems are notable because their thermal (and mechanical) responses tend to be much larger than those of inorganic solids, and because of their potential to provide insights into new mechanisms of thermal expansion. Metal-organic frameworks are also a challenge for first-principles DFT, not only because their unit cells typically contain a large number of atoms, but also because dispersion interactions dominate bonding in many materials in this family. However, methodological advances and rapid increases in computing power are putting these systems within reach.

\section{Acknowledgements}
This work was supported by the National Science Foundation. E. T. R. and N. A. B. were supported by DMR-1550347. Computational resources were provided by the Cornell Center for Advanced Computing and the Extreme Science and Engineering Discovery Environment (XSEDE) through allocation DMR-160052.

\bibliography{citations.bib}% Produces the bibliography via BibTeX.

\end{document}